\documentclass[
 reprint, aps, prx, 
superscriptaddress,
 amsmath,amssymb,
 aps, physrev,
]{revtex4-2}

\usepackage{graphicx}
\usepackage{dcolumn}
\usepackage{bm}
\usepackage[colorlinks=True,citecolor=blue,linkcolor=blue,urlcolor=blue]{hyperref}

\newcommand       \ba           {\begin{eqnarray}}
\newcommand       \ea           {\end{eqnarray}}
\begin{document}



\title{\textbf{Modeling transport in weakly collisional plasmas using thermodynamic forcing} }%

\author{Prakriti Pal Choudhury}
\email{prakriti.palchoudhury@physics.ox.ac.uk}
\affiliation{Department of Physics, University of Oxford, Parks Road, OX1 3PU, United Kingdom}
 
\author{Archie F. A. Bott}%
\affiliation{Department of Physics, University of Oxford, Parks Road, OX1 3PU, United Kingdom}


\begin{abstract}
 How momentum, energy, and magnetic fields are transported in the presence of macroscopic gradients is a fundamental question in plasma physics. Answering this question is especially challenging for weakly collisional, magnetized plasmas, where macroscopic gradients influence the plasma's microphysical structure. In this paper, we introduce thermodynamic forcing, a new method for systematically modeling how macroscopic gradients in magnetized or unmagnetized plasmas shape the distribution functions of constituent particles. In this method, we propose to apply an anomalous force to those particles inducing the anisotropy that would naturally emerge due to macroscopic gradients in weakly collisional plasmas {\color{black} in which thermal pressure is much larger than magnetic pressure}. We implement thermodynamic forcing in particle-in-cell (TF-PIC) simulations using a modified Vay particle pusher and validate it against analytic solutions of the equations of motion. We then carry out a series of simulations of electron-proton plasmas with periodic boundary conditions using TF-PIC. First, we confirm that the properties of two electron-scale kinetic instabilities -- one driven by a temperature gradient and the other by {\color{black} bulk-velocity gradient} -- are consistent with previous results. Then, we demonstrate that in the presence of {\color{black} both} macroscopic gradients, heat-flux saturation {\color{black}is mediated by the bulk-velocity-gradient-driven electron firehose instability rather than the temperature-gradient-driven whistler instability. This suggests that saturation mechanisms may differ from our current understanding in the presence of multiple free energy sources}. This work enables, for the first time, systematic and self-consistent transport modeling in weakly collisional plasmas, with broad applications in astrophysics, laser-plasma physics, and inertial confinement fusion.
\end{abstract}

\maketitle


\section{\label{sec:intro}Introduction} 

Plasma found in both astrophysical systems and high-energy laser experiments is often dilute, hot, and magnetized. The Coulomb mean free paths $\lambda_s$ of the plasma's constituent charged particles are {\color{black} small but not infinitesimal} fractions of the length scale $L$ that characterizes the macroscopic dynamics of the plasma. However, these mean free paths typically exceed the Larmor radii $\rho_s$ at which particles gyrate around magnetic field lines. This hierarchy of characteristic length scales means that these weakly collisional plasmas exhibit behavior that is fundamentally different from their strongly collisional counterparts.

In astrophysics, a significant fraction of the Universe's baryonic matter \cite{1992MNRAS_persic} surrounds the largest galaxies in the form of weakly collisional plasma: the so-called intracluster medium (ICM) of galaxy clusters.  As the largest gravitationally bound objects in the Universe, clusters are widely studied in astrophysics, galaxy formation and evolution, and cosmology \cite{1999ApJ_mohr, kravtsov2012formation}. They have also been observed extensively over many decades via X-rays and, more recently, the Sunyaev-Zeldovich effect \cite{fabian2000chandra, mulchaey2000x, rosati2002evolution,ettori2013mass, walker2019physics}. The ICM's low-temperature counterpart, the circumgalactic medium (CGM), is also a weakly collisional, magnetized plasma, and is detectable across multiple wavelengths \cite{tumlinson2017circumgalactic, faucher2023key}. 

In laser-plasma physics, coronal blow-off plasmas created by laser irradiation of solid targets are typically weakly collisional, and magnetic fields generated by the Biermann-battery mechanism are strong enough to make the Larmor radius $\rho_e$ of thermal electrons smaller than their Coulomb mean free path $\lambda_e$~\cite{Stamper_1971,Li_2007,Schoeffler_2016}. The hotspots of burning inertial-confinement-fusion (ICF) capsules, such as those recently created on the National Ignition Facility, are also weakly collisional~\cite{AbuShawareb_2024}, and recent simulations suggest that magnetic fields spontaneously generated by a variety of mechanisms are strong enough to cause $\rho_e \ll \lambda_e$~\cite{Walsh_2017,Sadler_2022}. 

Accurate modeling of heat and momentum transport is essential for addressing key problems in these physical systems. For example, in ICM physics, a classic problem is that of ``cooling flows" identified more than two decades ago~\cite{fabian1994cooling}: in the absence of heating, cluster cores should lose all their thermal energy due to {\color{black} thermal} bremsstrahlung radiation over a small fraction of the Hubble time, triggering a catastrophic collapse of the atmosphere. This catastrophe is not observed by the X-ray telescopes (e.g., \emph {Chandra, XMM-Newton}), which detect galaxy clusters routinely. There is observational evidence on the sources of heat in the ICM/CGM, such as active galactic nuclei and supernovae, but it is still unclear how energy is transported and redistributed from such spatially confined sources to the entire medium across a large distance, and prevent this catastrophe. A further complication arises due to persistence of sharp temperature contrasts in such atmospheres as seen in high resolution X-ray imaging. Thermal conduction is therefore widely discussed in galaxy cluster physics in the context of energy redistribution \cite{2003ApJ_zakamska, 2004MNRAS_voigt}. Thermal conduction may also play a significant role in the dynamics of accretion disks at low accretion rates and weak collisionality, such as those around supermassive black holes at the centers of galaxies -- including our own \cite{2007ApJ_johnson, 2007ApJ_sharma, 2016MNRAS_foucart}. In ICF research, heat conduction from the core of the hotspot to the surrounding dense plasma is thought to be the dominant loss mechanism immediately before ignition~\cite{AbuShawareb_2022}. 


Mounting empirical evidence suggests that classical theories of transport processes can often break down in weakly collisional, magnetized plasmas. Such theories assume that transport is mediated solely by Coulomb collisions and that the Coulomb mean free paths of particles are much smaller than macroscopic length scales ($\lambda_{e}, \lambda_i \ll L$). Further, they suggest that when $\rho_s \ll \lambda_s$, transport along magnetic field lines resembles that in unmagnetized plasma, while being suppressed across them. However, in the ICM core, where $\lambda_e \sim 10^{-2} L$, astronomical observations indicate that heat conduction is suppressed by several orders of magnitude relative to the classical Spitzer value \cite{1962pfig.book_spitzer}, to the extent that local, tangled, and stochastic magnetic field cannot account for suppression \cite{braginskii1965transport}. Recent experiments in which a turbulent, magnetized, weakly collisional plasma was created by the collision of laser-plasma jets also observed suppressed thermal conductivity \cite{2022SA_Meinecke}. In addition, X-ray observations of turbulence suggest that the effective viscosity in the ICM is suppressed by several orders of magnitude compared to the Braginskii viscosity \cite{2019NatAs_zhuravleva}. 

\begin{figure}
\centering
    \includegraphics[width=9cm]{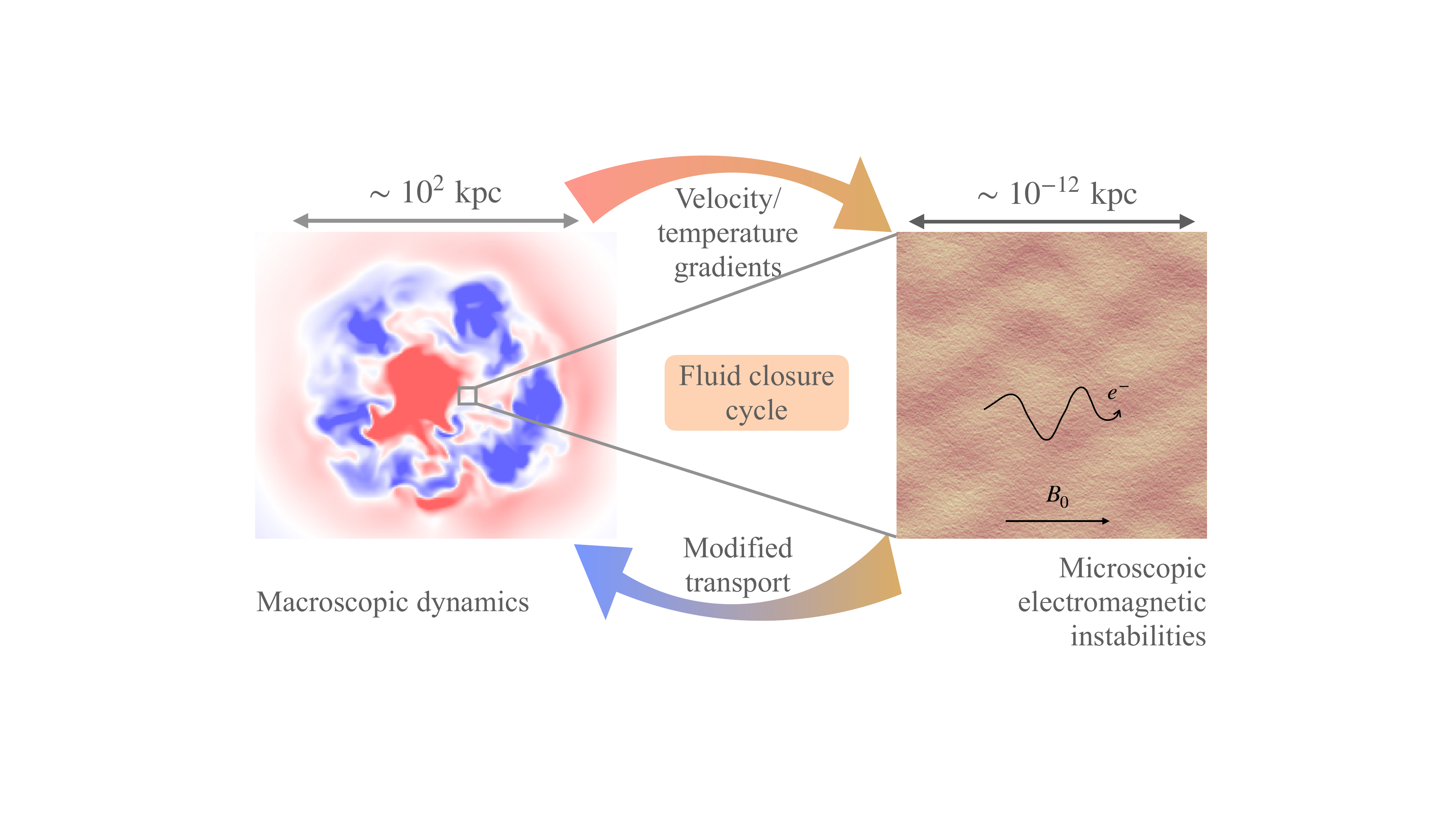}
    \centering
    \caption{Schematic diagram of self-consistent transport modeling in weakly collisional plasmas. The left panel shows the bulk velocities in a simulation of a hydrodynamic galaxy cluster~\cite{choudhury2022acoustic}, while the right panel shows the magnetic field of a kinetically unstable plasma mode, amplified via a kinetic instability, in a thermodynamically forced particle-in-cell simulation {\color{black} and the wavy arrow indicates that electron trajectories are influenced by the fluctuations in magnetic field}. In effect, the bulk flows are sources of free energy that drive kinetic plasma instabilities. These instabilities interact with the electrons and ions and affect the particle distribution, which in turn modifies macroscopic momentum and energy fluxes.}
    \label{fig:fig1}
\end{figure}
{\color{black} A} plausible explanation for suppressed transport is \emph{kinetic plasma instabilities}.
It is well established that collisionless plasmas with anisotropic distribution functions are generically susceptible to a range of kinetic instabilities. In plasmas with macroscopic gradients, streaming of particles from one region to another induces anisotropy. Some of these kinetic instabilities are suppressed when the anisotropy is regulated by collisionality, while others are stabilized if there is a macroscopic magnetic field whose energy is comparable to the thermal energy. However, in high-$\beta$ plasmas ($\beta \equiv 8 \pi p/B^2$, $p$ is thermal pressure of the plasma and $B$ is the amplitude of magnetic field threaded by the plasma), where the thermal energy greatly exceeds the magnetic energy, certain kinetic instabilities can be triggered by even a small anisotropy. Once destabilized, microscale electromagnetic fields can grow to levels sufficient to scatter particles, thereby enhancing the effective collisionality of the plasma. 
  
Over the last few decades, there has been a concerted effort to study the effect of kinetic instabilities on transport in plasmas. Studies of temperature gradients in collisionless plasmas using particle-in-cell (PIC) simulations have explored the role of the heat-flux driven whistler instability in suppressing heat transport along the magnetic field \cite{levinson1992,roberg2016suppression, komarov2016suppression, yerger2024collisionless}. The key finding of this research is that electromagnetic fluctuations can scatter particles, predominantly in their pitch angle, leading to the parallel electron heat flux $q_{\|} \sim q_{\rm fs}/\beta$ being suppressed by a factor of $\beta$ compared to the free-streaming level $q_{\rm fs}$. The details of the scattering process remain under debate \cite{ma2024whistler}. 
Another class of kinetic instabilities extensively studied in relation to momentum transport (or effective viscosity) in these plasmas includes the firehose and mirror instabilities \cite{PhysRev_parker, 1958RSPSA_chandra, schekochihin2008nonlinear, camporeale2008electron, kunz2014firehose, chen2016multi}. These instabilities are triggered by a reduction (or enhancement) in the perpendicular pressure $p_{\perp}$ relative to the parallel pressure $p_{\|}$, which in turn arises in the presence of bulk-velocity gradients at global scales. Both instabilities can cause electrons and ions to regulate the anisotropy $\Delta = p_{\perp}/p_{\|} - 1 \sim \beta^{-1}$ to small values scaling as $\beta^{-1}$, in contrast to the behavior of a double-adiabatic fluid~\cite{chew1956boltzmann}. PIC simulations have shown that the anisotropy is pinned at the marginal stability condition, leading to a non-isotropic distribution function that supports a suppressed momentum flux, compared to classical predictions. 

A key conclusion from these previous works is the necessity of a comprehensive understanding of kinetic instabilities, including their onset, saturation, and their impact on the distribution of particles, for modeling transport in weakly collisional plasmas, incorporating both Coulomb collisions and anomalous scattering from kinetic microinstabilities \cite{drake2021whistler}. However, a generalized solution to this problem has remained elusive. A major challenge is that, in weakly collisional plasmas, multiple kinetic instabilities driven by velocity and temperature gradients may coexist, depending on their dynamical length and time scales \cite{schwartz1980plasma,2024JPP_bott}. Quantifying the particle distribution at the saturated stage of multiple instabilities remains challenging. For instance, tentative evidence suggests that mirror modes also suppress thermal conduction \cite{2016MNRAS_komarov}. Multiple concurrent kinetic instabilities may mediate cooling and heating via thermal coupling between ions and electrons \cite{2023ApJ_zhdankin, ley2023heating}, which is crucial for understanding heating processes in astrophysical plasmas. Another major challenge is the computational cost of simulating increasingly large numerical domains, which is required for modeling transport with macroscopic-microscopic scale separations representative of astrophysical systems, while simultaneously resolving the Debye length.  

In this work, we address this challenge, presenting a new method -- \emph{thermodynamic forcing} -- that enables systematic modeling of transport in weakly collisional plasmas {\color{black} with $\beta \gg 1$}. This method thereby allows for the possibility of weakly collisional plasma systems such as the ICM to be modeled as fluids, even when classical transport theory fails. A schematic of our approach is shown in Fig.~\ref{fig:fig1}. {\color{black} The primary aim of this work is to introduce, formalize, and validate this method, rather than to extract new transport coefficients or identify previously unknown instabilities. Accordingly, the examples and simulations we present are chosen to establish the correctness, flexibility, and limitations of the thermodynamic forcing method by reproducing and extending well-studied results in a unified setting.
 It is worth clarifying that the term ``thermodynamic forcing" is distinct from the ``entropic force" that also arises in non-equilibrium statistical mechanics. The former refers to an externally imposed drive that produces velocity- or momentum-space anisotropy by injecting free energy into the system, for example through large-scale ($\gg \lambda_e$) gradients in temperature or flow. By contrast, an entropic force is an effective force that emerges internally as a system evolves toward a macrostate of higher entropy.}   

 The interplay between macroscopic and microscopic scales in such a system can be understood as follows. Macroscopic dynamics, such as those demonstrated in the hydrodynamic simulation shown in the left panel of Fig.~\ref{fig:fig1}, produce velocity and temperature gradients, along with variations in large-scale magnetic fields. These global dynamics act as sources of free energy, producing velocity-space anisotropy. This anisotropy drives kinetic instabilities, modifying particle distribution functions at saturation which, in turn, modifies global fluxes, and, consequently, the macroscopic evolution. Using PIC simulations, we demonstrate that the thermodynamic forcing method {\color{black} faithfully reproduces and generalizes} multiple sources and geometries of free energy, and thus enable comprehensive plasma transport modeling. This is in contrast to previous work on this problem, which has been specialized to particular problems. We further claim that the thermodynamic forcing method, including Coulomb collisions, has the potential to characterize anomalous scattering across a wide range of collisionality, accessing the transitional regime between classical and non-classical transport. The thermodynamic forcing method is straightforward to implement: by construction, it applies to periodic domains with homogeneous boundary conditions, while allowing complete control of the degree of anisotropy.  Our derivation of thermodynamic forcing also holds in weakly collisional, {\color{black} mildly} relativistic plasma, allowing us to treat the small fraction of suprathermal particles that are present in PIC simulations accurately, where electron temperatures are {\color{black} not many orders of magnitude smaller than} the rest mass energy. Thus, thermodynamic forcing provides a powerful and general approach for studying transport in weakly collisional plasmas across a wide range of physical conditions.
 
 The structure of our paper is as follows. We first establish the theoretical foundations of thermodynamic forcing (section  \ref{sec:model}), before demonstrating its implementation in test particle (section \ref{sec:singpart}) and fully kinetic simulations (section \ref{sec:sims}). We then discuss its implications for transport modeling across a range of physical regimes. Appendices \ref{app0}-\ref{app6} provide additional details and results that are important for our problem. 

\section{\label{sec:model}Modeling transport using thermodynamic forcing: theory}

We begin this section by outlining the conceptual basis for thermodynamic forcing as an approach for modeling the effect of macroscopic gradients on the distribution function, and thereby for determining plasma transport (section \ref{sec:concept}). In section \ref{sec:VDFderivation}, we then provide a formal derivation for thermodynamic forcing in one particular scenario: that of an electron temperature gradient driving an electron heat flux. Finally, we derive the specific form of thermodynamic forcing in non-relativistic plasma (\ref{sec:genmethod}) and in a plasma with relativistic electrons (\ref{sec:covmethod}). 

\subsection{Conceptual basis of thermodynamic forcing}
\label{sec:concept}

The core goal of any transport theory is to determine how macroscopic gradients in properties of a plasma -- for example, temperature or velocity -- give rise, microphysically, to fluxes of momentum or heat. In general terms, this relationship proceeds as follows. As soon as a macroscopic gradient develops in a plasma, anisotropy of the distribution arises due to the streaming of particles from one part of the plasma with a certain distribution to a second part with another. In the presence of collisionality -- which can either be Coulomb collisionality, or the effective collisionality arising due to the interaction of particles with destabilized plasma waves -- the streaming of particles and hence the distribution-function anisotropy is regulated. In a plasma where the mean free path $\lambda_\mathrm{mfp}$ associated with {\color{black} either the Coulomb collisionality or the effective collisionality} is much smaller than the macroscopic gradient scale $L$ (in other words, the plasma is either strongly or weakly collisional), this regulation occurs on a much greater timescale than that over which the plasma evolves macroscopically, and so a quasi-steady state of the distribution function is obtained in which its anisotropy is small: $(f_s-f_{\mathrm{M}s})/f_{\mathrm{M}s} \sim \lambda_\mathrm{mfp}/L \ll 1$ {\color{black} where $f_s$ is the particle distribution function of species $s$ and $f_{\mathrm{M}s}$ is the Maxwellian for the same species with spatially varying fluid variables}. This quasi-steady-state anisotropy, in turn, supports fluxes. In short, the effect of a macroscopic gradient microphysically is to generate distribution-function anisotropy, while collisionality acts to regulate it; the balance between these two physical processes then determines heat and momentum fluxes. 

The key idea that underlies thermodynamic forcing is that {\color{black} small} anisotropies $f_s-f_{\mathrm{M}s}$ of the distribution function that arise in a {\color{black} plasma with macroscopic gradients, due to the streaming of particles along those gradients, can be emulated \emph{within a homogeneous kinetic calculation} on scales much smaller than $L$}. This is achieved by adding a force on all particles that is spatially uniform but is dependent on the particles' velocity. From the perspective of individual particles, the two systems are not equivalent: the thermodynamic forcing accelerates (or decelerates) particles whose individual velocity would otherwise not change. However, an appropriately chosen thermodynamic force can modify the distribution function of those particles in approximately the same manner that particle streaming along a macroscopic gradient would, provided that the {\color{black} resulting} distribution-function anisotropy remains small. This latter condition arises because the error associated with the approximation is $\textit{O}(f_s/f_{\mathrm{M}s}-1)$. To guarantee the accuracy of the approximation, it is therefore necessary that the plasma's collisionality (Coulomb or effective) must ensure that $(f_s-f_{\mathrm{M}s})/f_{\mathrm{M}s} \sim \lambda_\mathrm{mfp}/L \ll 1$ for thermodynamic forcing to be applicable. {\color{black} In particular, thermodynamic forcing cannot be employed in collisionless plasmas, where $\lambda_\mathrm{mfp} \gtrsim L$.}

Thermodynamic forcing comes into its own when considering situations in which effective collisionality is generated by plasma kinetic instabilities. In scenarios where Coulomb collisionality is dominant, thermodynamic forcing is not strictly necessary because the collision operator in this scenario is known. The collision operator is independent of the macrosocopic dynamics, and the transport coefficients can be computed directly via the Chapman-Enskog expansion. However, in weakly collisional plasmas, thermodynamic forcing provides a method to model the distribution-function anisotropies that drive kinetic instabilities, and thereby allow the effective collisionality associated with those instabilities to evolve naturally. {\color{black} We note that thermodynamic forcing cannot be used to simulate instabilities whose linear physics explicitly requires spatial inhomogeneity of macroscopic plasma properties within the simulation domain (e.g. drift-wave instabilities~\cite{Horton_1999}).}

Thermodynamic forcing as a tool for determining transport properties offers two unique advantages over other approaches in which physical inhomogeneities are actually included. First of these is the technique's generalizability: determining the correct thermodynamic forcing for arbitrary temperature gradients, flow profiles, and magnetic field orientations is not much more challenging than the special cases. The same cannot be said for simulations that create flow or temperature gradients via specialized boundary conditions (BCs) e.g., shearing-box BCs or thermal baths at some boundaries, but not others. Simulations using thermodynamic forcing, by contrast, can simply employ periodic BCs. The second advantage is the homogeneity of the plasma in which thermodynamic forcing is employed. Consequently, the domain-averaged estimates of heat and momentum fluxes can be accurately determined. This improves the uncertainty on the calculation of these quantities compared to those from  simulations which directly simulate inhomogeneities, where only a subsection of the simulation can be used for calculating the mean properties of the plasma. 

  \subsection{A formal derivation of thermodynamic forcing}
\label{sec:VDFderivation}

We now show that the distribution-function anisotropy driven by thermodynamic forcing can be approximately the same as that arises naturally due to macroscopic gradients {\color{black} in a non-relativistic plasma}. For the sake of simplicity, we consider a special case for this derivation: that of a plasma whose distribution functions start from spatially uniform Maxwellians: $f_s(\boldsymbol{r},\boldsymbol{v},{\color{black}t=0}) = \bar{f}_{\mathrm{M}s}(v)$ for all species $s$, where $\boldsymbol{r}$ is spatial position, $\boldsymbol{v}$ is particle velocity, $v$ is the particle speed, $t$ is time, \color{black} and
\begin{equation}
\bar{f}_{\mathrm{M}s}(v) \equiv \frac{\bar{n}_{s}}{\pi^{3/2}\bar{v}_{\mathrm{th}s}^3} \exp{\left(-\frac{v^2}{\bar{v}_{\mathrm{th}s}^2}\right)} , 
\end{equation}
for uniform number densities $\bar{n}_{s}$ and thermal velocities $\bar{v}_{\mathrm{th}s}$ of species $s$. \color{black} We then assume that the plasma begins to develop a macroscopic electron temperature gradient of scale $L_{T} \equiv (\nabla \log{T_e})^{-1}$ due to localized heating and/or cooling {\color{black} over a characteristic timescale $\tau_{\rm heat} \sim L_{\rm T}/\bar{v}_{\mathrm{th}e}$ (we use a convention of negative temperature gradient)}. 

In response to this heating, the distribution functions $f_s$ evolve according to the kinetic equations
\ba
\nonumber
\frac{\partial f_{s}}{\partial t} &+& \boldsymbol{v} \cdot \boldsymbol{\nabla}{f_{s}} + \frac{Z_{s} e}{m_{s}} \Big( \boldsymbol{E} + \frac{\boldsymbol{v} \times \boldsymbol{B}}{c}\Big) \cdot \frac{\partial f_{s}}{\partial \boldsymbol{v}} \\
&=& \sum_{s^\prime} \mathcal{C}_{\rm c} (f_{s}, f_{s^\prime})  , \label{derivation_fullkinetic_eq}
\ea
where $Z_s$ is the charge of species $s$, $e$ the elementary charge, $m_s$ the mass of species $s$, $\boldsymbol{E}$ is the electric field, $\boldsymbol{B}$ the magnetic field, $c$ the speed of light, {\color{black} and} $\mathcal{C}_{\rm c} (f_{s}, f_{s^\prime})$ is the operator quantifying the effect of binary Coulomb collisions between species $s$ and $s^\prime$. For our purposes here, we do not need to specify any particular form of the collision operator. 

To rewrite this into a form that includes thermodynamic forcing explicitly, we first define a spatial averaging operator $\langle \cdot \rangle_l$ that takes an average over some spatial scale $l$ that is intermediate between the macroscopic and microscopic length scales in this problem: $L_{\rm T} \gg l \gg \lambda_{e} \sim \rho_e$, where we have assumed that the electron mean-free path $\lambda_{e}$ and electron Larmor radius $\rho_e$ are the same order. We then use this average to split the distribution functions and fields into macroscale and microscale components: 
\ba
f_{s} = \tilde{f}_{s} + \delta f_s , \; \boldsymbol{E} = \tilde{\boldsymbol{E}} + \delta \boldsymbol{E} ,\; \boldsymbol{B} = \tilde{\boldsymbol{B}} + \delta \boldsymbol{B} , 
\ea
where 
\ba
\langle f_{s} \rangle_l = \tilde{f}_{s} , \; \langle \boldsymbol{E} \rangle_l = \tilde{\boldsymbol{E}} , \; \langle \boldsymbol{B} \rangle_l = \tilde{\boldsymbol{B}}  ,  
\ea
and 
\ba
\langle \delta f_{s} \rangle_l = 0 , \; \langle \delta \boldsymbol{E} \rangle_l = 0 , \; \langle \delta \boldsymbol{B} \rangle_l = 0 .  
\ea
Note that we denote a quantity $X$ which varies in space over macroscopic scales via $\tilde{X}$, whereas uniform quantities are denoted by $\bar{X}$. The former implies an inhomogeneous medium at global scales (for example, due to the global temperature gradient), while the latter implies a homogeneous plasma at global scales (for example, a periodic box of plasma with uniform temperature, {\color{black} at scales smaller than the system's temperature gradient scale}). In what follows, we describe the method to produce the same anisotropies in the homogeneous plasma that happens in real inhomogeneous plasmas in astrophysics.

We then consider the evolution of the distribution functions on a timescale $t \sim \nu_{e}^{-1} \sim \beta_e^{-1} L_{T}/v_{\mathrm{th}e}$, where $\beta_e \sim L_{T}/\lambda_e \gg 1$ is the ratio of the electron thermal pressure to magnetic pressure {\color{black} (assumed large)} and {\color{black} $\nu_{e}$ is the Coulomb collision frequency}. {\color{black} It is on this timescale, which is assumed to be much smaller than the characteristic heating timescale $\tau_{\rm heat} \sim \beta_e \nu_e^{-1}$, that significant anisotropy of the distribution is expected to develop.} On this timescale, temperature equilibration between ions and electrons due to Coulomb collisions is negligible compared to the change in electron temperature ($t/\tau_{ei}^{\epsilon} \sim m_e/m_i \ll 1$ where $\tau_{ei}^{\epsilon}$ is the electron-ion temperature equilibration timescale), so the macroscale distribution of the ions remains unchanged from its initial distribution on this order of approximation. Consequently, only the electron distribution function will develop significant anisotropy due to the temperature gradient, and we can therefore specialize to the electron kinetic equation only [(\ref{derivation_fullkinetic_eq}), with $s = e$]. 

Next, we apply the spatial averaging operator to (\ref{derivation_fullkinetic_eq}) to obtain the evolution of $\tilde{f}_e$, 
\ba
\nonumber
\frac{\partial \tilde{f}_{e}}{\partial t} &+& \boldsymbol{v} \cdot \boldsymbol{\nabla}{\tilde{f}_{e}} - \frac{e}{m_{e}} \left( \tilde{\boldsymbol{E}} + \frac{\boldsymbol{v} \times \tilde{\boldsymbol{B}}}{c}\right) \cdot \frac{\partial \tilde{f}_{e}}{\partial \boldsymbol{v}} \\
&=& \sum_{s^\prime} \langle \mathcal{C}_{\rm c} (f_{e}, f_{s^\prime}) \rangle_l \nonumber \\ 
&& {\color{black}  +  \frac{e}{m_{e}} \left \langle \left( \delta \boldsymbol{E} + \frac{\boldsymbol{v} \times \delta \boldsymbol{B}}{c}\right) \cdot \frac{\partial \delta {f}_{e}}{\partial \boldsymbol{v}} \right \rangle_l } \; , \label{derivation_meankinetic_eq}
\ea 
where {\color{black} the last term on the right hand side of (\ref{derivation_meankinetic_eq}) models the effect on the macroscopic distribution function of particle scattering by electromagnetic waves}. An evolution equation for $\delta f_e$ can be obtained by taking the difference between (\ref{derivation_meankinetic_eq}) and (\ref{derivation_fullkinetic_eq}): 
\ba
\nonumber
\frac{\partial \delta {f}_{e}}{\partial t} &+& \boldsymbol{v} \cdot \boldsymbol{\nabla}{\delta {f}_{e}} - \frac{e}{m_{e}} \left( \tilde{\boldsymbol{E}} + \frac{\boldsymbol{v} \times \tilde{\boldsymbol{B}}}{c}\right) \cdot \frac{\partial \delta {f}_{e}}{\partial \boldsymbol{v}} \nonumber \\
&& - \frac{e}{m_{e}} \left( \delta {\boldsymbol{E}} + \frac{\boldsymbol{v} \times \delta {\boldsymbol{B}}}{c}\right) \cdot \frac{\partial \tilde{f}_{e}}{\partial \boldsymbol{v}} \nonumber \\
&=& \frac{e}{m_{e}} \left( \delta \boldsymbol{E} + \frac{\boldsymbol{v} \times \delta \boldsymbol{B}}{c}\right) \cdot \frac{\partial \delta {f}_{e}}{\partial \boldsymbol{v}} \nonumber \\ 
&& {\color{black}  -  \frac{e}{m_{e}} \left \langle \left( \delta \boldsymbol{E} + \frac{\boldsymbol{v} \times \delta \boldsymbol{B}}{c}\right) \cdot \frac{\partial \delta {f}_{e}}{\partial \boldsymbol{v}} \right \rangle_l } \nonumber \\ && + \sum_{s^\prime} \left[\mathcal{C_{\rm c}} (f_{e}, f_{s^\prime})  - \langle \mathcal{C}_{\rm c} (f_{e}, f_{s^\prime}) \rangle_l \right] . \label{derivation_diffkinetic_eq}
\ea 

We then adopt the following ordering of quantities with respect to each other,
\begin{equation}
\frac{\delta f_e}{\tilde{f}_e} \sim \frac{c}{v_{\mathrm{th}e}} \frac{|\tilde{\boldsymbol{E}}|}{|\tilde{\boldsymbol{B}}|}\sim \frac{\lambda_e}{L_{\rm T}} \frac{|\delta \boldsymbol{B}|}{|\tilde{\boldsymbol{B}}|} \sim \frac{c}{v_{\mathrm{th}e}} \frac{|\delta \boldsymbol{E}|}{|\tilde{\boldsymbol{B}}|} \sim \frac{\lambda_e}{L_{\rm T}} \ll 1 , \label{ordering}
\end{equation}
and assume that the wavenumber $k$ of microscale physics satisfies $k \rho_e \sim 1$. This latter assumption is the most appropriate one, because the kinetic instabilities of significance that can be triggered in this scenario arise at 
electron Larmor scales. Expanding the macroscopic distribution function in ${\lambda_e}/{L_{\rm T}} \ll 1$,
\ba
\tilde{f}_{e} = \tilde{f}^{(0)}_{e} + \tilde{f}^{(1)}_{e} + ... \, , 
\ea
we can now solve {\color{black} the evolution equation (\ref{derivation_meankinetic_eq}) for $\tilde{f}_e$} order by order. 

To leading order, we find
\begin{equation}
\frac{e \boldsymbol{v} \times \tilde{\boldsymbol{B}}}{m_e c} \cdot \frac{\partial \tilde{f}_{e}^{(0)}}{\partial \boldsymbol{v}}
 + {\color{black}\mathcal{C}_{\rm c} (\tilde{f}_{e}^{(0)}, \tilde{f}_{e}^{(0)}) 
 +  \mathcal{C}_{\rm c} (\tilde{f}_{e}^{(0)}, \bar{f}_{\mathrm{Mi}}) } = 0 . \label{derivation_meankinetic_eq_0}
\end{equation}
 {\color{black} We note that the electromagnetic scattering term in (\ref{derivation_meankinetic_eq}) is higher order under the ordering (\ref{ordering}) because it is proportional to $\delta f_e$. By contrast, the Lorentz term due to the mean magnetic field acts on the leading-order distribution $\tilde{f}_{e}^{(0)}$. Eq. (\ref{derivation_meankinetic_eq_0}) is identical to the leading-order kinetic equation appearing in Braginskii's derivation of classical transport theory, for which a Maxwellian whose characteristic thermal speed can vary in space over the macroscopic scale length $L_{\rm T}$, is the unique solution~\cite{helander2005collisional}: 
 \ba
\tilde{f}^{(0)}_{e} = \tilde{f}_{\mathrm{M}e} \, . \label{derivation_dist_0}
\ea}

To first order in ${\lambda_e}/{L_{\rm T}} \ll 1$, (\ref{derivation_meankinetic_eq}) becomes
\ba
\frac{\partial \tilde{f}_{e}^{(1)}}{\partial t} & - & \frac{e\boldsymbol{v} \times \tilde{\boldsymbol{B}}}{m_e c} \cdot \frac{\partial \tilde{f}_{e}^{(1)}}{\partial \boldsymbol{v}} \nonumber \\ & {\color{black} -} & {\color{black} \frac{e}{m_{e}} \left \langle  \frac{\boldsymbol{v} \times \delta \boldsymbol{B}}{c} \cdot \frac{\partial \delta {f}_{e}}{\partial \boldsymbol{v}} \right \rangle_l   - \mathcal{C}_{\rm c} (\tilde{f}_{e}^{(1)}, \bar{f}_{\mathrm{M}e}) } \nonumber \\ & \color{black} - & {\color{black} \mathcal{C}_{\rm c} (\bar{f}_{\mathrm{M}e},\tilde{f}_{e}^{(1)})  
  - \mathcal{C}_{\rm c} (\tilde{f}_{e}^{(1)}, \bar{f}_{\mathrm{Mi}}) } = \mathcal{S}_e, \label{derivation_meankinetic_eq_1}
\ea
where 
\begin{equation}
\mathcal{S}_e = - \frac{\partial \tilde{f}_{\mathrm{M}e}}{\partial t} - \boldsymbol{v} \cdot \boldsymbol{\nabla} \tilde{f}_{\mathrm{M}e} + \frac{e}{m_{e}} \boldsymbol{\tilde{E}} \cdot \frac{\partial \tilde{f}_{\mathrm{M}e}}{\partial \boldsymbol{v}}    \label{derivation_sourceterm}
\end{equation}
can be interpreted physically as the source of the distribution function's anisotropy that arises due to the macroscopic gradients of (\ref{derivation_dist_0}). {\color{black} We note that the time derivative of $\tilde{f}_e^{(1)}$ is included here at the same order as that of $\tilde{f}_e^{(0)}$. This is because our system begins in a state of thermodynamic equilibrium (i.e., $\tilde{f}_e^{(1)}(t = 0) = 0$), and develops its first-order correction $\tilde{f}_e^{(1)} \sim  (\lambda_e/L_{\rm T})\tilde{f}_{\mathrm{M}e}$ on the collisional timescale $t \sim \nu_e^{-1}$. Estimating the time derivative, it follows that 
\begin{equation}
\frac{\partial \tilde{f}^{(1)}_e}{\partial t} \sim \frac{\tilde{f}_e^{(1)}}{t} \sim \frac{\lambda_e}{L_{\rm T}} \nu_e \tilde{f}_{\mathrm{M}e} \sim \frac{\tilde{f}_{\mathrm{M}e}}{\tau_{\rm heat}} \sim \frac{\partial \tilde{f}^{(0)}_e}{\partial t}, 
\end{equation}
justifying its inclusion.}

Evaluating (\ref{derivation_sourceterm}) explicitly, we find that
\begin{equation}
\mathcal{S}_e = {\color{black}-}\tilde{f}_{\mathrm{M}e} \left(\frac{{v}^2}{v^2_{\mathrm{th}e}} - \frac{5}{2}\right)\boldsymbol{v}\cdot {\boldsymbol \nabla}\ln T_e = -\frac{1}{m_e} \frac{\partial}{\partial \boldsymbol{v}} \boldsymbol \cdot \left(\boldsymbol{F}_{\rm T} \bar{f}_{\mathrm{M}e}\right) , \label{derivation_sourceterm_simp}
\end{equation}
where 
\begin{equation}
\boldsymbol{F}_{\rm T} \equiv -\frac{m_{e}}{2}\left(v^2 - \frac{3}{2}v^2_{\mathrm{ th}e}\right) {\boldsymbol \nabla}\ln T_e
\end{equation}
is the thermodynamic force, and the second equality in (\ref{derivation_sourceterm_simp}) holds to the order of the asymptotic approximation (i.e. we have neglected terms of order $\lambda_e^2/L_{\rm T}^2$). Now, defining the modified, macroscopic distribution function 
\begin{equation}
\bar{f}_e \equiv \bar{f}_{\mathrm{M}e} + f_e^{(1)} ,
\end{equation}
it follows that
\ba
\frac{\partial \bar{f}_{e}}{\partial t} & - & \frac{e\boldsymbol{v} \times \bar{\boldsymbol{B}}}{m_e c} \cdot \frac{\partial \bar{f}_{e}}{\partial \boldsymbol{v}} + \frac{1}{m_e} \frac{\partial}{\partial \boldsymbol{v}} \boldsymbol \cdot \left(\boldsymbol{F}_{\rm T} \bar{f}_{e}\right) \nonumber \\
& \color{black} = & \color{black} \mathcal{C}_{\rm c} (\bar{f}_{e}, \bar{f}_{e}) 
 +  \mathcal{C}_{\rm c} (\bar{f}_{e}, \bar{f}_{\mathrm{Mi}}) \nonumber \\ 
&& {\color{black}  +  \frac{e}{m_{e}} \left \langle \left( \delta \boldsymbol{E} + \frac{\boldsymbol{v} \times \delta \boldsymbol{B}}{c}\right) \cdot \frac{\partial \delta {f}_{e}}{\partial \boldsymbol{v}} \right \rangle_l } \; . \label{derivation_meankinetic_eq_moddist}
\ea
 Here, $\bar{\boldsymbol{B}} \equiv \tilde{\boldsymbol{B}}(t = 0)$ is a spatially uniform magnetic field, and we have again neglected terms of order $\lambda_e^2/L_{\rm T}^2$. 

Now considering again the evolution equation for $\delta f_e$ [cf. (\ref{derivation_diffkinetic_eq})], we find that, to leading order in $\lambda_e/L_{\rm T} \ll 1$,
\ba
\nonumber
\frac{\partial \delta {f}_{e}}{\partial t} &+& \boldsymbol{v} \cdot \boldsymbol{\nabla}{\delta {f}_{e}} - \frac{e \boldsymbol{v} \times \bar{\boldsymbol{B}}}{m_e c} \cdot \frac{\partial \delta {f}_{e}}{\partial \boldsymbol{v}} \nonumber \\
&& - \frac{e}{m_{e}} \left( \delta {\boldsymbol{E}} + \frac{\boldsymbol{v} \times \delta {\boldsymbol{B}}}{c}\right) \cdot \frac{\partial \bar{f}_{e}}{\partial \boldsymbol{v}} \nonumber \\
&=& \frac{e}{m_{e}} \left( \delta \boldsymbol{E} + \frac{\boldsymbol{v} \times \delta \boldsymbol{B}}{c}\right) \cdot \frac{\partial \delta {f}_{e}}{\partial \boldsymbol{v}} \nonumber \\ 
&& {\color{black}  -  \frac{e}{m_{e}} \left \langle \left( \delta \boldsymbol{E} + \frac{\boldsymbol{v} \times \delta \boldsymbol{B}}{c}\right) \cdot \frac{\partial \delta {f}_{e}}{\partial \boldsymbol{v}} \right \rangle_l } \nonumber \\ && + \;  \mathcal{C_{\rm c}} (\delta f_{e}, \bar{f}_{e}) + \mathcal{C_{\rm c}} (\bar{f}_{e},{\color{black} \delta} {f}_{e}) + \mathcal{C_{\rm c}} (\delta f_{e}, \bar{f}_{\mathrm{M}i}). \quad \quad \label{derivation_diffkinetic_eq_moddist}
\ea 
Finally, we define modified distribution function
\begin{equation}
f_e^{*} \equiv \bar{f}_{e} + \delta f_e ,
\end{equation}
which we emphasize is uniform in space to the order of the approximation. 
Adding together (\ref{derivation_meankinetic_eq_moddist}) and (\ref{derivation_diffkinetic_eq_moddist})
we conclude that $f_e^{*}$ satisfies the kinetic equation
\ba
\nonumber
\frac{\partial f_{e}^{*}}{\partial t} &+& \boldsymbol{v} \cdot \boldsymbol{\nabla}{f_{e}^{*}} + \frac{1}{m_e} \frac{\partial}{\partial \boldsymbol{v}} \boldsymbol \cdot \left(\boldsymbol{F}_{\rm T} {f}_{e}^{*}\right) \nonumber \\
& - &\frac{e}{m_{e}} \left[\delta \boldsymbol{E} + \frac{\boldsymbol{v} \times \left(\bar{\boldsymbol{B}}+\delta \boldsymbol{B}\right)}{c}\right] \cdot \frac{\partial f_{e}^{*}}{\partial \boldsymbol{v}} \nonumber \\
&=& \mathcal{C}_{\rm c} (f_{e}^{*}, f_{e}^{*}) +\mathcal{C}_{\rm c} (f_{e}^{*}, f_{\mathrm{M}i}) . \label{derivation_fullkinetic_eq_moddist}
\ea
 The derivation is complete: we have shown that we can determine the anisotropy $f_e^{(1)} = f_e-\tilde{f}_{\mathrm{M}e}$ of the true distribution function by solving (\ref{derivation_fullkinetic_eq_moddist}), which includes thermodynamic forcing, for $f_e^{*}$, and relating it to $f_e^{(1)}$ via $f_e^{(1)} = f_e^{*}-\bar{f}_{\mathrm{M}e}$. 

Formal derivations for other types of thermodynamic forcing follow similarly, though for the sake of this paper's readibility we do not repeat them here. Instead, in the next section, we describe a simple method by which the most general form of thermodynamic forcing can be deduced. 

\subsection{General form of thermodynamic forcing: non-relativistic case}
\label{sec:genmethod}

To determine an approach of finding the correct thermodynamic forcing $\boldsymbol{F}_{\rm T}$ in a (non-relativistic) plasma that supports arbitrary temperature and velocity gradients, we reconsider its derivation in section \ref{sec:VDFderivation}. $\boldsymbol{F}_{\rm T}$ was calculated from its relation (\ref{derivation_sourceterm_simp}) to the inhomogeneous term $\mathcal{S}_e$ that drives anisotropy in the kinetic equation (\ref{derivation_meankinetic_eq_1}) to first order; in turn, $\mathcal{S}_e$ was evaluated by substituting the zeroth-order macroscopic distribution function $\tilde{f}_e^{(0)}$ -- which is Maxwellian -- into the full kinetic equation (\ref{derivation_meankinetic_eq}) for $\tilde{f}_e$, with the microscale physics playing no direct role. Therefore, if we simply  neglect all microscale fluctuations, and substitute 
\begin{equation}
{f}_s = \tilde{f}_{\mathrm{M}s} + {f}_s^{(1)} \label{dist_func_Maxsub}
\end{equation}
into the full kinetic equations, respectively, we can determine $\mathcal{S}_s$ and therefore $\boldsymbol{F}_{\rm T}$ by grouping all the inhomogeneous terms. 

 We start from the general kinetic equation (\ref{derivation_fullkinetic_eq}), and, now assuming each species $s$ has a bulk fluid velocity $\boldsymbol{V}_s$, move to a new coordinate frame $(\boldsymbol{r},\boldsymbol{v},t) \rightarrow (\boldsymbol{r},\boldsymbol{v}^\prime_s,t)$, where the peculiar velocity $\boldsymbol{v}^\prime_s$ is given by $\boldsymbol{v}^\prime_s = \boldsymbol{v} - \boldsymbol{V}_{s}$. Under this transformation, the kinetic equation becomes
\ba
\nonumber
\sum_{ s^\prime} \mathcal{C}_{\rm c} (f_{s}, f_{ s^\prime}) &-& \frac{Z_{s} e}{m_{s} c} (\boldsymbol{v}^\prime_s \times \boldsymbol{B}) \cdot \frac{\partial f_{s}}{\partial \boldsymbol{v^\prime_s}} \nonumber \\ &=& \frac{\mathrm{D} f_{s}}{\mathrm{D}t} + \boldsymbol{v}^\prime_s \cdot \boldsymbol{\nabla} f_{s} - \boldsymbol{v}^\prime_s \cdot (\boldsymbol{\nabla}\boldsymbol{V}_{s})\cdot \frac{\partial f_{s}}{\partial \boldsymbol{v}^\prime_s} \nonumber \\
&& + \Big( \frac{Z_{s} e}{m_{s}} \boldsymbol{E}^\prime - \frac{\mathrm{D} \boldsymbol{V}_s}{\mathrm{D}t} \Big)\cdot \frac{\partial f_{s}}{\partial \boldsymbol{v}^\prime_s}
, \label{force_fullkineticinframe_eq}
\ea 
where
\begin{equation}
\frac{\mathrm{D}}{\mathrm{D}t} = \frac{\partial}{\partial t} + \boldsymbol{V}_{s}\cdot \boldsymbol {\nabla}
\end{equation}
is the convective derivative, and $\boldsymbol{E}^\prime = \boldsymbol{E} + (\boldsymbol{V}_{s} \times \boldsymbol{B})/c$ is the electric field in a frame co-moving with the fluid. We arrange the terms in (\ref{force_fullkineticinframe_eq}) in such a way that the inhomogeneous terms arise on the right-hand side. Because of the distinct structure of the collision operator in the kinetic equation for the electron distribution function versus the ion distribution function, we now consider these two cases separately. 

First, for the electrons, the electron-ion collision operator $\mathcal{C}_{\mathrm{c}}(f_e,f_i)$ can be approximated as 
\ba
\mathcal{C}_{\mathrm{c}}(f_e,f_i) &=& \mathcal{C}_{ei}^{(0)}(f_e) +  \mathcal{C}_{ei}^{(1)}(f_e) , 
\ea
where $\mathcal{C}_{ei}^{(0)}(f_e)$ is a pitch-angle scattering operator, and $\mathcal{C}_{ei}^{(1)}(f_e)$ a drag operator that is smaller than $\mathcal{C}_{ei}^{(0)}$ by a factor $\sim \sqrt{m_e/m_i}$. Both terms are independent of the ion distribution function. We then use the substitution (\ref{dist_func_Maxsub}) in (\ref{force_fullkineticinframe_eq}), and combine terms by order in $\lambda_e/L \sim \sqrt{m_e/m_i}\ll 1$. The zeroth-order equation vanishes; the first-order equation for electrons is 
 \ba
   \mathcal{C}_{\rm c}(f_{e}^{(1)}, \tilde{f}_{{\rm M}e}) &+& \mathcal{C}_{\rm c}(\tilde{f}_{{\rm M}e},f_{e}^{(1)}) + \mathcal{C}^{(0)}_{ei}(f_{e}^{(1)}) \nonumber \\
   & - & \frac{Z_{e} e}{m_{e} c} ({\boldsymbol{v}^\prime_e \times \boldsymbol{B}}) \cdot \frac{\partial f_{e}^{(1)}}{\partial \boldsymbol{v}^\prime_e} \nonumber \\ 
   & = & \frac{\mathrm{D} f_{\mathrm{M}e}}{\mathrm{D}t} + \boldsymbol{v}^\prime_e \cdot \boldsymbol{\nabla} f_{\mathrm{M}e}  - \boldsymbol{v}^\prime_e \cdot (\boldsymbol{\nabla}\boldsymbol{V}_{e})\cdot \frac{\partial {f}_{\mathrm{M}e}}{\partial \boldsymbol{v}^\prime_e} \nonumber \\
   && + \left( \frac{Z_{e} e}{m_{e}} \boldsymbol{E}^\prime - \frac{\mathrm{D} \boldsymbol{V}_{e}}{\mathrm{D}t} \right)\cdot \frac{\partial f_{\mathrm{M}e}}{\partial \boldsymbol{v}^\prime_e} \nonumber \\
   && - \; \mathcal{C}^{(1)}_{ei} (\tilde{f}_{\mathrm{M}e}) . 
   \label{eq: ce_pair}
 \ea
 Now calculating the right hand side using a spatially varying Maxwellian, we find that 
 \ba
  \mathcal{C}_{\rm c}(f_{e}^{(1)}, \tilde{f}_{{\rm M}e}) & + & \mathcal{C}_{\rm c}(\tilde{f}_{{\rm M}e},f_{e}^{(1)})+ \mathcal{C}^{(0)}_{ei}(f_{e}^{(1)}) \nonumber \\ &-& \frac{Z_{e} e}{m_{e} c} ({\boldsymbol{v}^\prime_e \times \boldsymbol{B}}) \cdot \frac{\partial f_{e}^{(1)}}{\partial \boldsymbol{v}^\prime_e} = - \mathcal{S}_e , \label{eq:source_eq_elec}
 \ea
 where
\ba 
\label{eq:S}
\mathcal{S}_e &\equiv& -f_{\mathrm{M}e} \left[ \left(\frac{{v^\prime_e}^2}{v^2_{\mathrm{th} e}} \right. \right.  -  \left. \left. \frac{5}{2}\right)\boldsymbol{v}^\prime_e\cdot {\boldsymbol \nabla}\ln T_{e} \right. \nonumber \\
&& \left. + \frac{\boldsymbol{v}^\prime_e\cdot \boldsymbol{\mathcal{R}}_{e}}{p_{e}} + \frac{m_e \nu_{ei}(v')\boldsymbol{v}^\prime_e\cdot\boldsymbol{u}_{ei}}{T_{e}} \right. \nonumber \\
&& \left. + \frac{m_{e}}{2T_{e}} \boldsymbol{v}^\prime_e\cdot \boldsymbol{\mathcal{W}}_e \cdot \boldsymbol{v}^\prime_e \right], \label{eq:electron_source_term_full}  
\ea
{\color{black}and $Z_e=-1$}.
 Here, $\boldsymbol{u}_{ei}$ is the difference in bulk-velocity between the electrons and ions,  
 \ba 
\label{eq:W}
\boldsymbol{\mathcal{W}}_s = \boldsymbol{\nabla}\boldsymbol{V}_{s} + \left(\boldsymbol{\nabla}\boldsymbol{V}_{s}\right)^{\rm T} - \frac{2}{3} \left(\boldsymbol{\nabla}\cdot \boldsymbol{V}_{s}\right) \boldsymbol{I}
   \ea
is the (traceless) rate-of-strain tensor of the fluid motions of species $s$,
\ba
\label{eq:R}
\boldsymbol{\mathcal{R}}_{e} = \int m_{e}\boldsymbol{v}^\prime_e \mathcal{C}_{\rm c} (f_{e}, f_{i}) \mathrm{d}^3 \boldsymbol{v}^\prime_s
\ea
is the frictional force on the electrons due to collisions with the ions, and 
\ba
\nu_{ei}(v^\prime_e) = \frac{3\sqrt{\pi}}{4\tau_{e}} \left( \frac{v^3_{{\rm th}e}}{v^{\prime 3}_e}\right)
\ea
is the velocity dependent collision frequency. We note that the third term of the right hand side of (\ref{eq:electron_source_term_full}) has an apparent singularity as $v'$ tends to zero; this is, in practice, balanced by a second singularity in the electron-ion collision operator.

Next, it is a simple calculation to show that the terms introducing inhomogeneities in the homogeneous plasma can be written as 
\begin{equation}
\mathcal{S}_e = -\frac{1}{m_{e}}\frac{\partial}{\partial \boldsymbol{v}^\prime_e}\cdot (\boldsymbol{F}_{\mathrm{T}e} f_{\mathrm{M}e} ), 
\end{equation}
where $\boldsymbol{F}_{\mathrm{T}e}$ is the thermodynamic force on electrons. Evaluating the exact form of $\boldsymbol{F}_{\mathrm{T}e}$, we find that
\ba
\boldsymbol{F}_{\mathrm{T}e} &=& -\frac{m_{e}}{2}\left[ \left(v^{\prime 2}_e - \frac{3}{2}v^2_{{\rm th}e}\right) \frac{\boldsymbol{\hat{a}}}{L_{\rm T}}  + \boldsymbol{\mathcal{W}}_e\cdot \boldsymbol{v}^\prime_e  \right. \nonumber \\
&& \left. + \frac{\boldsymbol{\mathcal{R}}_{e}}{{\color{black}m_e}n_{e}}+ \mathcal{F}_{\rm drift} \hat{\boldsymbol{u}}_{ei}\right] \label{ft_force_norel}
\ea
where $\boldsymbol{\hat{a}}$ is the direction of the temperature gradient,
\ba
\mathcal{F}_{\rm drift} = \frac{3\sqrt{\pi}v_{{\rm th}e}}{\tau_e} \left[ \frac{1}{v^\prime_e} {\color{black}-} \frac{\sqrt{\pi} e^{{v^{\prime 2}}/{v^2_{{\rm th}e}}}{\rm erfc}{\left({v^\prime_e}/{v_{{\rm th}e}}\right)}}{v_{{\rm th}e}}\right]  \:
\label{eq:driftforce}
\ea
and $\hat{\boldsymbol{u}}_{ei}$ is the direction of the electron-ion drift.  
  
  The thermodynamic force on electrons defined by (\ref{ft_force_norel}) has several components. The first of these drives the form of distribution-function anisotropy that arises when the plasma supports a macroscopic electron temperature gradient. The second component drives the anisotropy caused by macroscopic gradients in the bulk flow of the electrons. The third force denotes frictional force and the fourth is due to the Coulomb collisional drift. We note that the collisional drift force diverges at small electron velocities -- which, as previously discussed, is an effect that should be balanced by enhanced Coulomb collisionality at small velocities.  The frictional force and collisional drift are not known to cause any kinetic instabilities~\cite{2024JPP_bott}, and so, for simplicity's sake, we chose to neglect the friction and collisional drift force in our subsequent calculations. 

The calculation of the thermodynamic forcing on the ions is similar to that of the electrons, with one important difference: ion-electron collisions are negligible compared to ion-collisions, and so the analogue of (\ref{eq:source_eq_elec}) is 
 \ba
  && \mathcal{C}_{\rm c}(f_{i}^{(1)}, \tilde{f}_{{\rm M}i}) + \mathcal{C}_{\rm c}( \tilde{f}_{{\rm M}i},f_{i}^{(1)}) \nonumber \\
 && \quad - \frac{Z_{i} {\color{black}e}}{m_{i} c} ({\boldsymbol{v}^\prime_i \times \boldsymbol{B}}) \cdot \frac{\partial f_{i}^{(1)}}{\partial \boldsymbol{v}^\prime_i} = \mathcal{S}_i , \label{eq:source_eq_ion}
 \ea
 where
\ba 
\label{eq:Si}
\mathcal{S}_i &\equiv& -f_{\mathrm{M}i} \left[ \left(\frac{{v^\prime_i}^2}{v^2_{\mathrm{th} i}} \right. \right.  -  \left. \left. \frac{5}{2}\right)\boldsymbol{v}^\prime_i\cdot {\boldsymbol \nabla}\ln T_{i} \right. \nonumber \\
&& \left. + \frac{m_{i}}{2T_{i}} \boldsymbol{v}^\prime_i\cdot \boldsymbol{\mathcal{W}}_i \cdot \boldsymbol{v}^\prime_i \right]. \label{eq:ion_source_term_full}  
\ea
It follows that the thermodynamic force on the ions is

\ba
\boldsymbol{F}_{\mathrm{T}i} &=& -\frac{m_{i}}{2}\left[ \left(v^{\prime 2}_i - \frac{3}{2}v^2_{{\rm th}i}\right) \frac{\boldsymbol{\hat{a}}}{L_{\rm T}}  + \boldsymbol{\mathcal{W}}_i\cdot \boldsymbol{v}^\prime_i \right] \label{ft_force_ions_norel} .
\ea
In effect, ions are not subject to terms associated with the frictional force or collisional drift. 


\subsection{General form of thermodynamic forcing: {\color{black} weakly} relativistic case}
\label{sec:covmethod}

Although our focus in this paper is modeling heat and momentum transport in non-relativistic, weakly collisional plasmas, a key issue when running kinetic simulations of such plasmas is their considerable computational expense. To circumvent this, it is a standard practice when simulating non-relativistic plasmas to use  values of $v_{\mathrm{th}e}/c$ (the ratio of electron thermal speed to the speed of light) {\color{black}not much smaller than unity}, and reduced values of $m_{i}/m_{e}$ (the ratio of ion to electron mass), respectively, to reduce the characteristic evolution timescale of the plasma and hence reduce the computational cost of the simulations. However, if we wish to use a similar approach when simulating plasmas that are subject to thermodynamic forcing, we must determine a form of such forcing that is valid in plasmas in which some electrons have energies that are comparable to their relativistic rest mass. This is particularly pertinent because suprathermal electrons are thought to play a key role in heat transport -- so, if we are to use values of $v_{\mathrm{th}e}/c$ {\color{black} that are not much smaller than unity} in simulations, such particles will be relativistic. Furthermore, when particles are kicked individually by thermodynamic forces, it is plausible that some of them will be accelerated, attaining relativistic energies. Therefore, in this section we outline a relativistic calculation of the thermodynamic force $\boldsymbol{F}_{\mathrm{T}e}$ on electrons. 

The most general method for determining thermodynamic forcing in a fully relativistic plasma is to analyze the covariant form of the kinetic equation for its distribution functions, allowing for the possibility that the bulk-velocity $\boldsymbol{V}_s$ of plasma species $s$ could be comparable to the speed of light. However, because the plasmas most pertinent to our study are not undergoing relativistic bulk motions, allowing for relativistic motions of suprathermal electrons only ($\boldsymbol{V}_s \sim v_{\mathrm{th}i} \sim (m_e/m_i)^{1/2} v_{\mathrm{th}e} \ll v_{\mathrm{th}e} \ll c$) will be sufficient and also enables the tractability of calculating the thermodynamic forcing analytically. 

To calculate the thermodynamic forcing in this case, we start the calculation from {\color{black} the special relativistic electron kinetic equation (c.f., Eqs. (2.17) and (2.22) of \cite{2002cercignani})}:
\ba
\nonumber
 \frac{\partial f_{e}}{\partial t} &+& \gamma^{-1}_{\rm p}\boldsymbol{u} \cdot \boldsymbol{\nabla}{f_{e}} + \frac{1}{m_{e}}\frac{\partial}{\partial  \boldsymbol{u}}\cdot(f_{e} \boldsymbol{F}_{\rm Lorentz}) \\
 &=& \frac{1}{\gamma_{\rm p}}\left[{\mathcal{C}}_{\rm c,rel} (f_{e},f_e) +{\mathcal{C}}_{\rm c,rel} (f_{e},f_i) \right] , \quad 
 \label{eq:covform}
\ea
where $\gamma_{\rm p}={(1-{v^{2}}/{c^2})}^{-1/2}$ is the Lorentz boost associated with individual particles, {\color{black}$\boldsymbol{u}=\gamma_{\rm p}\boldsymbol{v}$ is the 3-velocity in special relativity}, and ${\mathcal{C}}_{\rm c,rel}$ is the collision operator associated with Coulomb collisions of relativistic particles (we do not need its explicit form for our purposes). Alternately, the Lorentz factor is $\gamma_{\rm p}={(1+u^2/c^2)}^{1/2}$.

Similarly to the non-relativistic case, we 
transform the coordinate frame to write the relativistic Vlasov equation (\ref{eq:covform}) in terms of the peculiar velocity as
$(\boldsymbol{r},\boldsymbol{u},t) \rightarrow (\boldsymbol{r},\boldsymbol{u}^\prime_s,t)$, where the peculiar 3-velocity $\boldsymbol{u}^\prime_s$ is now given by the identity (see Appendix \ref{app0}): 
\ba
\boldsymbol{u}^\prime_e & = & \boldsymbol{u} - \gamma^\prime_{\rm p} \boldsymbol{V}_e , 
\label{eq:rel_v_add}
\ea
where $\boldsymbol{u}^\prime_e = \gamma^\prime_{\rm p} \boldsymbol{v}^\prime_e$, $\gamma_{e}={(1-{|\boldsymbol{V}_e|^{2}}/{c^2})}^{-1/2}$ is the Lorentz boost associated with the bulk electron flow. Assuming that $|\boldsymbol{V}_e| \ll c$ ($\gamma_e\approx1$), it can be shown from (\ref{eq:rel_v_add}), for particles with $\gamma_p - 1 \lesssim 1$, the coordinate transform is given to $\textit{O}(V_e/c)$ accuracy by
\ba
\frac{\partial}{\partial t} & \rightarrow & \frac{\partial}{\partial t} - \gamma^{\prime}_{\rm p} \frac{\partial \boldsymbol{V}_s}{\partial t} \cdot 
\frac{\partial}{\partial \boldsymbol{u}^\prime_s} \quad , \\
\nabla & \rightarrow & \nabla - \gamma^{\prime}_{\rm p} \nabla \boldsymbol{V}_s \cdot 
\frac{\partial}{\partial \boldsymbol{u}^\prime_s} \quad , \\
\frac{\partial}{\partial \boldsymbol{u}} & \rightarrow & \frac{\partial}{\partial \boldsymbol{u}^\prime_s} \quad , 
\ea
where $\gamma_{\rm p}^{\prime} ={(1-{v^{\prime 2}_e}/{c^2})}^{-1/2} \approx \gamma_{\rm p}$. The electron kinetic equation is then
\ba
\nonumber
 \frac{\partial f_{e}}{\partial t} & +&  \gamma^{\prime -1}_{\rm p}\boldsymbol{u}^\prime_e \cdot \boldsymbol{\nabla}{f_{e}} - \left[\frac{e}{m_e} \left(\boldsymbol{E}' + \frac{\boldsymbol{u}^\prime_e \times \boldsymbol{B}}{\gamma^\prime_{\rm p}c}\right) \nonumber \right. \\
 &+& \left. \gamma_{\rm p}^{\prime} \frac{\mathrm{D} \boldsymbol{V}_s}{\mathrm{D} t} + \boldsymbol{u}^\prime_e \cdot \boldsymbol{\nabla} \boldsymbol{V}_e \right] \cdot\frac{\partial f_e}{\partial \boldsymbol{u}^\prime_e} \nonumber \\
 &=& \frac{1}{\gamma_{\rm p}'}\left[{\mathcal{C}}_{\rm c,rel} (f_{e},f_e) +{\mathcal{C}}_{\rm c,rel} (f_{e},f_i) \right] . 
 \label{eq:covform_pv}
\ea

We then adopt a similar approach to that employed in section (\ref{sec:genmethod}): we neglect microscale fluctuations, and suppose that the distribution function $f_e$ takes the form
\begin{equation}
{f}_e = {f}_{\mathrm{MJ}e} + {f}_e^{(1)} \label{dist_func_MJsub} . 
\end{equation}
Here, we have assumed that $f_e$ is the relativistic generalization of a Maxwellian -- a Maxwell-J\" uttner distribution function ${f}_{\mathrm{MJ}e}$ -- to zeroth order in $\lambda_e/L \ll 1$. The Maxwell-J\" uttner distribution function {\color{black} in the momentum space} is given by 
\ba
{f}_{\mathrm{MJ}e}(\boldsymbol{u}^{\prime}_e) &=& \frac{n_{e} e^{-{\gamma_{\rm p}^{\prime}}/{\theta_e}}}{4\pi m_{e}^3 c^3\theta_e K_2 (\theta_e^{-1})}
\ea
where $\theta_e = {k_{\rm B}T_{e}}/{m_{e} c^2}$, and $K_2(\alpha)$ is the modified Bessel function of second kind. 

Substituting (\ref{dist_func_MJsub}) into (\ref{eq:covform_pv}), and combining terms order by order in $\lambda_e/L \ll 1$, the zeroth-order equation again vanishes, while the first-order becomes
\ba
   &&\frac{1}{\gamma^{\prime}_{\rm p}} \left[ \mathcal{C}_{\rm c, rel}(f_{e}^{(1)}, {f}_{\mathrm{MJ}e}) + \mathcal{C}_{\rm c, rel}( {f}_{\mathrm{MJ}e},f_{e}^{(1)}) \right. \nonumber \\
   && \left. + \mathcal{C}^{(0)}_{ei,{\rm rel}}(f_{e}^{(1)}) \right]+ \frac{e}{m_{e} \gamma^{\prime}_{\rm p} c} ({\boldsymbol{u}^\prime_e \times \boldsymbol{B}}) \cdot \frac{\partial f_{e}^{(1)}}{\partial \boldsymbol{u}^\prime_e}  \nonumber \\
   &= & \frac{\mathrm{D} {f}_{\mathrm{MJ}e}}{\mathrm{D}t} + \gamma^{\prime -1}_{\rm p}\boldsymbol{u}^\prime_e \cdot \boldsymbol{\nabla} {f}_{\mathrm{MJ}e} - \boldsymbol{u}^\prime_e \cdot (\boldsymbol{\nabla}\boldsymbol{V}_{e})\cdot \frac{\partial {f}_{\mathrm{MJ}e}}{\partial \boldsymbol{u}^\prime_e} \nonumber \\
   && - \left( \frac{e}{m_{e}} \boldsymbol{E}^\prime + \frac{\mathrm{D} \boldsymbol{V}_{e}}{\mathrm{D}t} \right)\cdot \frac{\partial {f}_{\mathrm{MJ}e}}{\partial \boldsymbol{u}^\prime_e} . 
   \label{eq: ce_pair_rel}
 \ea
Here, the electron-ion collision operator is approximately a pitch-angle scattering operator: $\mathcal{C}_{\rm c, rel} (f_{e},f_i) \approx \mathcal{C}^{(0)}_{ei, {\rm rel}}(f_{e})$. For moderately relativistic electrons colliding with ions, the drag operator $\mathcal{C}^{(1)}_{ei, {\rm rel}} (f_{e})$ is $\textit{O}(m_e/m_i)$ compared to $\mathcal{C}^{(0)}_{ei, {\rm rel}}(f_{e})$, and therefore does not appear in the first-order equation. 

We now evaluate the right-hand side of (\ref{eq: ce_pair_rel}) using the identity
\ba
\frac{\partial f_{{\rm MJ}e}}{\partial \boldsymbol{u}^\prime_e}=\frac{1}{\gamma^{\prime3}_{\rm p}}\frac{\partial f_{{\rm MJ}e} }{\partial \boldsymbol{v}^\prime_e} = \frac{\boldsymbol{v}^\prime_e}{c^2}\frac{\partial f_{{\rm MJ}e}}{\partial \gamma_{\rm p}} = - \frac{\boldsymbol{v}^\prime_e}{\theta_e c^2}f_{{\rm MJ}e} , 
\ea
and also
\ba
\frac{\mathrm{D}}{\mathrm{D}t} {f}_{\mathrm{MJ}e} & \approx &  \frac{\mathrm{D} \ln n_{e}}{\mathrm{D}t} + \frac{\mathrm{D} \ln T_{e}}{\mathrm{D}t} \left( \frac{\gamma_{\rm p}^\prime - 1}{\theta_e} - \frac{3}{2}\right) , \quad \\
\nabla {f}_{\mathrm{MJ}e} & \approx & \nabla \ln n_{e} + \nabla \ln T_{e}  \left( \frac{\gamma_{\rm p}^\prime - 1}{\theta_e} - \frac{3}{2}\right) , \quad
\ea
where we have performed a subsidiary expansion in $\theta_e \ll 1$ of the Bessel function of the second kind and its derivative  [${K^{\prime}_2(\theta_e^{-1})}/{K_2(\theta_e^{-1})}\approx -1+\theta_e/2$].  
We conclude that 
\ba
   &&\frac{1}{\gamma^{\prime}_{\rm p}} \left[ \mathcal{C}_{\rm c, rel}(f_{e}^{(1)}, {f}_{\mathrm{MJ}e}) + \mathcal{C}_{\rm c, rel}( {f}_{\mathrm{MJ}e},f_{e}^{(1)}) + \mathcal{C}^{(0)}_{ei, {\rm rel}}(f_{e}^{(1)}) \right]\nonumber \\
   &&  + \frac{e}{m_{e} \gamma^\prime_{\rm p} c} ({\boldsymbol{u}^\prime_e \times \boldsymbol{B}}) \cdot \frac{\partial f_{e}^{(1)}}{\partial \boldsymbol{u}^\prime_e} = -\mathcal{S}_{e} , 
   \label{eq: ce_pair_rel_simp}
 \ea
where the relativistic analogue to the source term (\ref{eq:S}) is
\ba
\nonumber
\mathcal{S}_{e} &=& -f_{{\rm MJ}e}\left[ \boldsymbol{v}^{\prime}_e \cdot \boldsymbol{\nabla}\ln T_{e}\Big(\frac{\gamma_{\rm p}^{\prime} - 1}{\theta_e} - \frac{5}{2} \Big)\right. \\ 
\nonumber
&& \left.+ \frac{\gamma_{\rm p}^{\prime}}{2\theta_e c^2}\boldsymbol{v}^\prime_e\cdot \boldsymbol{\mathcal{W}_e} \cdot \boldsymbol{v}^\prime_e \right. \\ 
&& \left. + \boldsymbol{\nabla}\cdot\boldsymbol{V}_{e} \left(-\frac{2}{3}\frac{\gamma_{\rm p}^{\prime} - 1}{\theta_e} + \frac{1}{3} \frac{\gamma_{\rm p}^{\prime} - 1/\gamma_{\rm p}^{\prime}}{\theta_e}\right)  \right] . 
\ea
The first and second terms on the right hand side are the (straight-forward) relativistic generalisations of free-energy sources driven by temperature and bulk-velocity gradients, respectively. The third term, by contrast, does not appear in our non-relativistic calculation; this is because, in the non-relativistic limit $v \ll c$, this term is $\textit{O}(v^2/c^2)$ compared to the other terms. 

Analogously to the non-relativistic case, it can be shown that
\begin{equation}
\mathcal{S}_e = -\frac{1}{m_{e}}\frac{\partial}{\partial \boldsymbol{u}^\prime_e}\cdot (\boldsymbol{F}_{\mathrm{T}e} f_{{\rm MJ}e} ). 
\end{equation}
where
\ba
\nonumber
\boldsymbol{F}_{{\rm T}e} &=& -m_e \left[ \frac{1}{2} \frac{ v^2_{{\rm th}e}}{L_{\rm T}} \Big( \frac{\gamma_{\rm p}^{\prime} -1}{\theta_e} -\frac{3}{2} \Big) \hat{\boldsymbol{a}} + \frac{1} {2}\boldsymbol{\mathcal{W}}_e\cdot \boldsymbol{u}^\prime_e \right. \\
&& \left. \qquad + \frac{\boldsymbol{\nabla}\cdot \boldsymbol{V}_{e}}{3}  \boldsymbol{u}^\prime_e \left(\frac{\gamma_{\rm p}^{\prime}-1}{\gamma_{\rm p}^{\prime} +1} \right)\right] \label{ft_force_rel}
\ea
is the thermodynamic force on moderately relativistic electrons. As expected, the first and second components of the thermodynamic force are straightforward generalizations of non-relativistic temperature gradient-driven and {\color{black} bulk-velocity gradient} driven thermodynamic forces [cf. (\ref{ft_force_norel})]. It is unknown whether the third component (see Appendix \ref{app1} for a derivation) of the thermodynamic force, only relevant for relativistic particles, can drive new kinetic instabilities. We will not explore this component of the thermodynamic force in detail moving forward in this work, since the dominant population of particles in our PIC simulations (discussed in section \ref{sec:sims}) are weakly relativistic.

\section{Numerical implementation of thermodynamic forcing}
\label{sec:singpart}
\begin{figure*}
    \includegraphics[width=18cm]{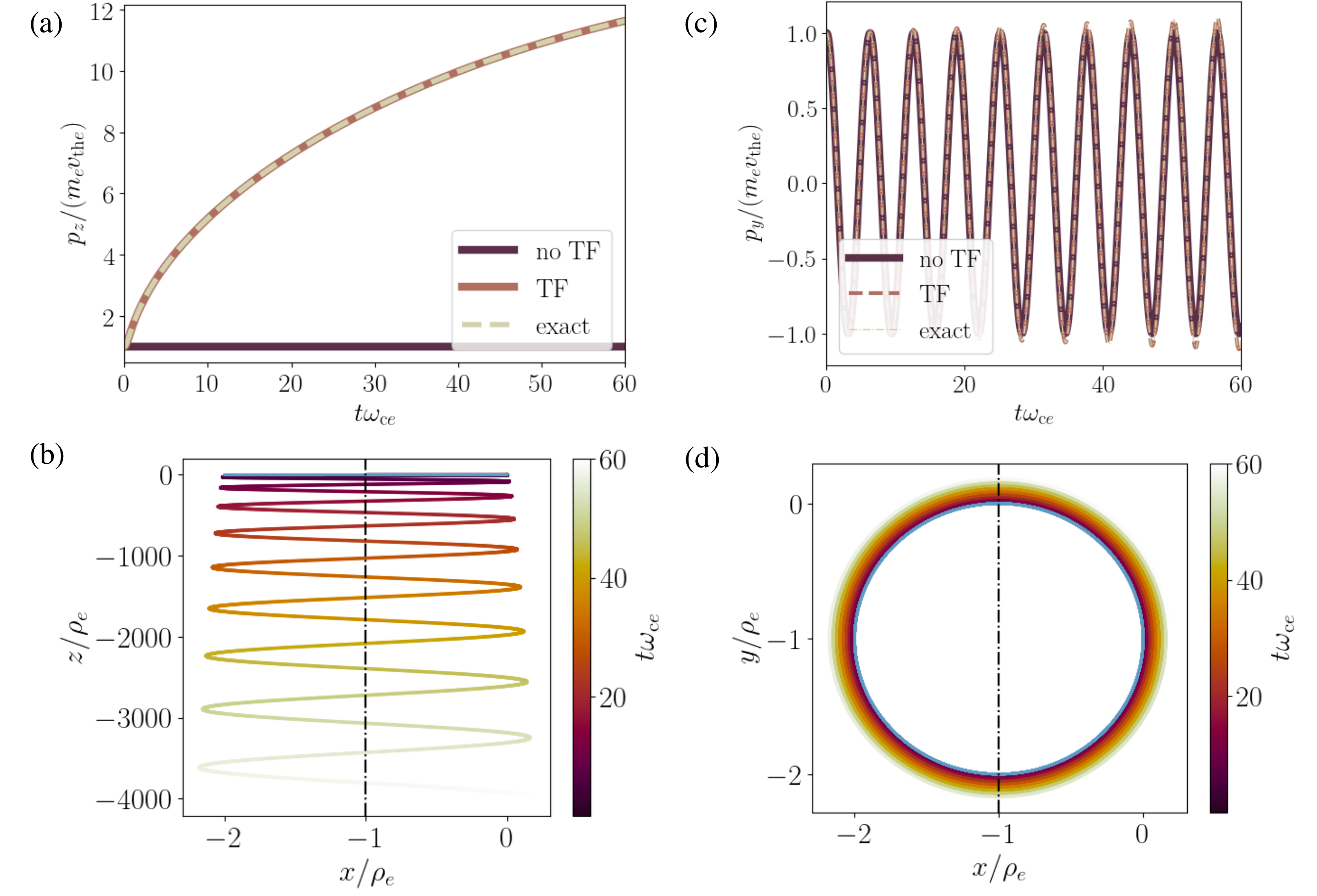}
    \centering
    \caption{(a) Time evolution of a particle's parallel (with respect to the magnetic field) momentum with and without TF (temperature-gradient) along with analytical prediction. (b) The spatial drift of the particle due to TF (temperature-gradient only) along the magnetic field. (c) The evolution of one component of the particle's perpendicular momentum with time and comparison with and without TF ({\color{black} bulk-velocity gradient} only) along with analytical prediction. (d) The trajectory of the particle in the perpendicular plane to the magnetic field with TF ({\color{black} bulk-velocity gradient}). The non-relativistic version is in Appendix \ref{app2}.}
    \label{fig:vay_velo}
\end{figure*}
 
Now that we have determined analytic expressions for thermodynamic forcing, we next consider its numerical implementation into kinetic simulations. This is a non-trivial step, because the thermodynamic forces given by (\ref{ft_force_norel}) and (\ref{ft_force_ions_norel}) for non-relativistic particles, and by (\ref{ft_force_rel}) for relativistic particles, are characteristically different from the (real) Lorentz forces that act on charged particles. The thermodynamic forces $\boldsymbol{F}_{{\rm T}s}$ are uniform in space, but in general depend on the particle velocity in quite a complicated way, which could, in principle, render standard numerical approaches unusable. For reasons of computational cost, we choose to implement the force in PIC simulations as opposed to a Vlasov-Fokker-Planck solver. In this section, we therefore consider how to implement thermodynamic forcing on (macro)particles (section \ref{sec:singpart_partpusher_ft}), then test this implementation on test particles (section \ref{sec:singpart_tests}). 

\subsection{Adding thermodynamic forcing to particle pushers} \label{sec:singpart_partpusher_ft}

To add thermodynamic forcing into PIC simulations, we need to implement $\boldsymbol{F}_{{\rm T}s}$ in the integrator step for the particles' equations of motion. The standard particle pusher algorithms applicable in non-relativistic and relativistic cases are the Boris pusher \cite{boris1970} and the Vay pusher \cite{vay2008simulation}. The former is a second-order leapfrog integrator with staggered discrete time points for velocity and position updates. This integrator conserves phase-space volume, produces errors in energy that are bounded, which implies good energy conservation, and is stable for sufficiently small time steps. Thus, for a single particle, a Boris pusher produces smooth energy-conserving orbits. Appendix \ref{app2} shows single particle tests of our force in the Boris pusher. However, for relativistic particles, the Boris integrator may produce large errors. Because a part of the electric field is converted to a magnetic field and vice versa under Lorentz transformations, the integrator must exactly cancel out the forces caused by the additional electric field in a new Lorentz frame by the equivalent additional magnetic field. In the Boris pusher, this does not happen. The Vay pusher provides a solution to this problem. In this section, we therefore describe the implementation of the relativistic form of $\boldsymbol{F}_{{\rm T}s}$ in the Vay pusher.

The components of the thermodynamic force (\ref{ft_force_rel}) depend on velocity in different ways. The first component -- the temperature-gradient-driven thermodynamic force -- only depends explicitly on the magnitude of the momentum (or velocity) of any particle, rather than its direction. For a given particle, dependence of the force on just $\gamma_p$ is equivalent to pushing the particle in a direction independent of its direction of motion. Thus, this can be implemented as an effective electric field,
\ba
\boldsymbol{E}_{\rm eff} \equiv \boldsymbol{E} &+&  \frac{1}{2} \frac{m_{s} v^2_{{\rm th}s}}{q_sL_{\rm T}} \Big( \frac{\gamma_{\rm p} -1}{\theta_s} -\frac{3}{2} \Big) \hat{\boldsymbol{a}} . 
\ea
 The second component -- the {\color{black} bulk-velocity gradient} driven thermodynamic force -- is dependent on the direction of the particle's motion. Hence, this force needs to be added as an operator-splitting step at the end of the pusher.
The Vay algorithm is therefore modified as follows {\color{black} to advance particles from ${t}_{i-1}$ to ${t}_i$, where ${t}_{i} \equiv i \; dt$ is the discrete simulation time, and $dt$ is the timestep:
\ba
\nonumber
\boldsymbol{p}^{*} &=& \boldsymbol{p}^{i-1} + {q_s {d}t} [\boldsymbol{E}_{{\rm eff}} (\gamma^{i-1}_{{\rm p}}) +{0.5 (\boldsymbol{v}^{{\it i}-{\rm 1}} \times \boldsymbol{B}^{{\it i}-{\rm 1}})}] , \\
\nonumber
\boldsymbol{p}^{i} &=& \frac{1}{1+|\boldsymbol{\upsilon}|^2}[\boldsymbol{p}^{*} + {(\boldsymbol{p}^{*}\cdot \boldsymbol{\upsilon})} \boldsymbol{\upsilon} + \boldsymbol{p}^{*} \times \boldsymbol{\upsilon}] , \\
\boldsymbol{p}^{i}_{\rm final} &=& \boldsymbol{\mathcal{J}}^{-1} \boldsymbol{p}^{i}
\label{eq:shcomp}
\ea
where $\boldsymbol{p} = m_e\gamma_{\rm p} \boldsymbol{v}$, 
\ba
\nonumber
\gamma^{i}_{{\rm p}} & =&  2^{-1/2}\sqrt{\sigma + \sqrt{\sigma^2 + 4|\boldsymbol{\tau}|^2 + w^2}} ,
\ea
 $\boldsymbol{\tau} = (q_s{d}t/2)\boldsymbol{B}$, $w = c^{-1}\boldsymbol{p}^{*}\cdot \boldsymbol{\tau}$, $\sigma = (\gamma^{\prime 2}_{\rm p} - |\boldsymbol{\tau}
 |^2)$, $\gamma^{\prime}_{{\rm p}}=\sqrt{1 + |\boldsymbol{p}^{*}|^2/m^2_ec^2}$, $\boldsymbol{\upsilon} = \boldsymbol{\tau}/\gamma^i_{{\rm p}}$, and
 \ba
 \boldsymbol{\mathcal{J}} \equiv \boldsymbol{I} - 0.5 { d}t\boldsymbol{\mathcal{W}} . \label{Jmat_def}
 \ea
 }The other aspects of the algorithm are identical to those presented in \cite{vay2008simulation}. {\color{black} The last step, given by (\ref{eq:shcomp}), updates the final momentum at time ${t}_i$ after adding the thermodynamic {\color{black} bulk-velocity gradient} force}. We note that for the algorithm to be usable, 
the matrix $\boldsymbol{\mathcal{J}}$ needs to be non-singular. 

 We use an implicit first-order method for the operator splitting step and rearrange the discretized equation related to $\boldsymbol{\mathcal{W}}$ to obtain the matrix algebraic equation (\ref{eq:shcomp}). Conceptually, any general shear/compression tensor can be modeled and explored by only parameterizing the tensor with a timescale for shear/compression/expansion, ${\color{black} \tau_{\rm comp}}$. 

\subsection{Tests of algorithm on single particles} \label{sec:singpart_tests}

We now test this numerical implementation of thermodynamic forcing on a single test particle in order to confirm that no numerical instabilities appear due to inclusion of this additional force. For this test, we chose to consider the motion of such particles in a static and uniform magnetic field. This scenario is favorable because of its analytical tractability, which means that we can compare our numerical solutions against the analytic ones. We implement the algorithm described in section \ref{sec:singpart_partpusher_ft} in a {\tt Python} code. 

For single particle dynamics, we use the Larmor radius, thermal velocity, and inverse Larmor frequency ($\rho_{e}, v_{{\rm th}e}, \omega^{-1}_{{\rm c}e}$) as normalisation units in our modified Vay algorithm. For a single particle, there is no intrinsic thermal velocity; but in $\boldsymbol{F}_{{\rm T}e}$ the presence of $v_{{\rm th}e}$ makes it a convenient normalization unit. A physically intuitive interpretation is to consider drawing a particle from the distribution with thermal temperature $\theta_e$. The fixed free parameters that we use for the tests with both forces are (in the above code units) $\boldsymbol{B}=\omega_{ {\rm c}e}q_e^{-1}\boldsymbol{\hat{z}}, c=300v_{{\rm th}e}, m_e=1,q_e=1, dt=.01\omega^{-1}_{{\rm c}e}, t_{\rm stop}=60\omega^{-1}_{{\rm c}e}$, and $\theta_s=k_{\rm B}T_{s}/m_sc^2= 0.1$. The particle completes many orbits during the {\color{black} elapsed time $t_{\rm stop}$}. The initial position is $\boldsymbol{{x}}=-\rho_e\boldsymbol{\hat{x}}$, and the initial velocity is $\boldsymbol{{v}}=v_{{\rm th}e}\boldsymbol{\hat{x}}$. 
 
\subsubsection{Test for temperature gradient force}
In this section, we demonstrate that a single particle evolves as predicted analytically when the temperature gradient force is treated an effective electric field. The equations of motion for the particle subject to the temperature-gradient driven force are  
\ba
\nonumber
\frac{d p_{\|}}{dt} &=& \frac{\theta_e c^2}{L_{\rm T}} \Big( \frac{\gamma_{\rm p} - 1}{\theta_e} - \frac{3}{2}\Big),\\
\frac{d p_{\perp}}{dt} &=& 0  . 
\ea
The analytical solution can be written in the form
\ba
\mathcal{F}_2 (\gamma_{\rm p}) - \mathcal{F}_2 (\gamma_{\rm p}(t=0)) &=& \frac{\theta_e c^2t}{L_{\rm T}}
\label{eq:sing_TG} , 
\ea
where $\mathcal{F}_2(\gamma_{\rm p})$ is a non-trivial function of the Lorentz factor (the analytical solution against which we compare the single particle trajectory is provided in Appendix \ref{app3}), and can be inverted to calculate ${p}_{\parallel}= m_s\gamma_{\rm p} {v}_{\parallel}$ at each time. 
For the single particle test, we use $L_{\rm T}=5000 \rho_e$. The parallel velocity evolution (upper left) and spatial drift (lower left) as functions of time are shown in Fig.~\ref{fig:vay_velo}. Parallel drift does not occur in the absence of the force, and we find very close agreement between the analytical and numerical results in the presence of TF. Thus, the temperature gradient force should gradually increase the heat flux of an initially thermal particle distribution along the parallel direction. 
\begin{figure*}
    \includegraphics[width=18cm]{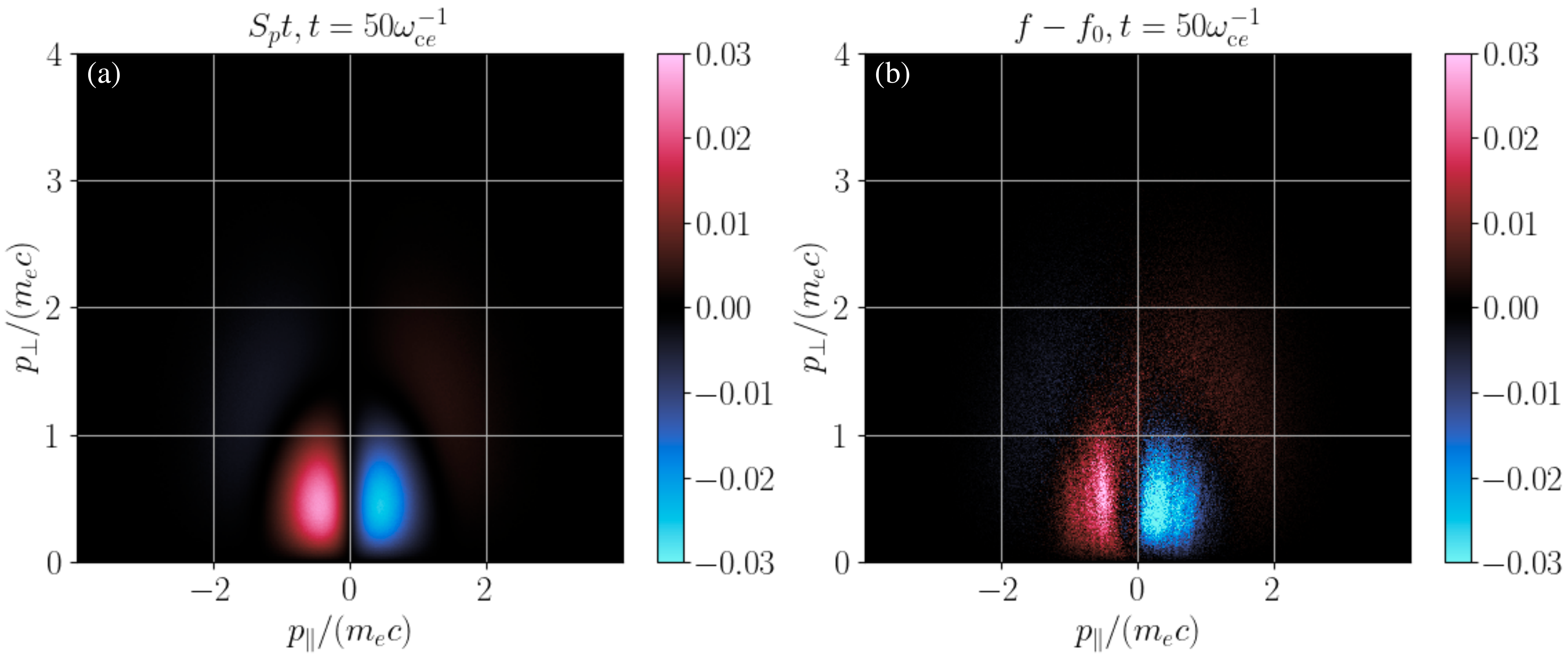}
    \centering
    \caption{Comparison between the (a) analytical ($S_{\rm p}t$), and (b) numerical ($f-f_0$) momentum space anisotropy in the parallel direction produced by TF (temperature-gradient), where $f_0$ and $f$ are the distribution functions at the initial time and a later time $t$, respectively, and $S_{\rm p}$ for a $\beta_e=60$ electron-proton plasma is given by (\ref{eq:Sp1}) in section \ref{sec:alignedgrad}, in two-dimensional momentum space. In the simulations, resonant dark lines are visible that deviate from straight lines around $p_{\parallel}/m_ec =1$, as expected for relativistic resonance; see (\ref{eq:reso_gen3}) in Appendix \ref{sec:resonance}.}
    \label{fig:wanisotropy}
\end{figure*}

One subtle issue may arise for suprathermal particles, which we identify from the non-relativistic and relativistic analytical solutions. In the non-relativistic case, the temperature-gradient-driven force may lead to an unbounded increase in the parallel velocity when it acts on a particle over a timescale $L_{\rm T}/{v}_{\rm r}$ where ${v}^2_{\rm r}/v^2_{{\rm th}e}={v}^2_{\perp}/v^2_{{\rm th}e} - 3/2$ (\ref{boris_tg}). Although $L_{\rm T}$ is large, this situation may arise dynamically for suprathermal particles in particle-in-cell (PIC) simulations. A similar situation arises in the relativistic case (\ref{vaytg}), where the initial $\boldsymbol{p}_{\perp}=m_e\gamma_{\rm p} \boldsymbol{v}_{\perp}$ sets the trajectory of the particle. We observe non-zero rates of change on this runaway timescale for a small fraction of electrons, specifically in the PIC simulation with lower $\beta_e$ discussed in section \ref{sec:alignedgrad}. We discuss this issue further in section \ref{sec:disc}.
\subsubsection{Test for the {\color{black} bulk-velocity gradient} force}
 For the {\color{black} bulk-velocity gradient} driven force, the analytical solution for the particle momentum satisfies $p_j\propto e^{\lambda_j {\color{black}t}}$ where the equation of motion for the test particle is
\ba
\nonumber
\frac{d p_i}{dt} &=& 0.5 \mathcal{W}_{ij} p_j + \epsilon_{ijk} v_j B_k 
\ea
The above equation can be solved as an eigenproblem, $\mathcal{M}_{ij}p_j=\lambda_jp_j$,  $\epsilon_{ijk}$ is the Levi-Civita function, $i,j,k$ run over the three components of a vector, and $\mathcal{M}_{ij} = \mathcal{W}_{ij} + \epsilon_{ijk} B_k \gamma^{-1}_{\rm p}$. The evolution of the perpendicular momentum components $\boldsymbol{p}_{\perp}$ must follow a growing or decaying oscillation according to the corresponding eigenvalue (oscillations due to the Lorentz force, growth or decay due to the {\color{black} bulk-velocity gradient} force), when the {\color{black} gradient} is in the perpendicular plane; therefore, we consider a bulk-velocity field
\ba
\boldsymbol{V}_{s} = \frac{x}{{\color{black} \tau_{\rm comp}}} \boldsymbol{\hat{x}} + \frac{y}{{\color{black} \tau_{\rm comp}}} \boldsymbol{\hat{y}} . \label{BV_field_perp}
\ea
 $\boldsymbol{\mathcal{W}}$ is a purely diagonal matrix for this choice of {\color{black} bulk-velocity field}. 
In the PIC simulations (section \ref{sec:sims}), we also use a similar $\boldsymbol{V}_{s}$ in the plane perpendicular to the guide field. We use ${\color{black} \tau_{\rm comp}} = 200\omega^{-1}_{{\rm c}e}$ in the single-particle test. In Fig.~\ref{fig:vay_velo}, we shown the evolution of one perpendicular velocity components (upper right) and the orbit in the perpendicular plane (lower right) as functions of time. Without TF, the test shows the gyrating oscillation, and with TF we find very close agreement between the numerical trajectory and analytical solution. The {\color{black} bulk-velocity gradient} force therefore gradually increases the temperature anisotropy of an initially thermal particle distribution.

\section{Thermodynamically forced (TF) PIC simulations}
\label{sec:sims}

In this section, we discuss the details of the computational set-up and two classes of microinstabilities (driven by the temperature gradient and {\color{black} bulk-velocity gradient} force, respectively) as well as the possibility of their joint occurrence. In order to carry out first-principle kinetic PIC simulations, we use {\tt OSIRIS} \cite{fonseca2002osiris}, a three-dimensional, fully relativistic, massively parallelized code. We use the Vay particle pusher scheme in the code, which we modify using the force implementation described in the previous section, implemented both as a net electric field and as a separate operator-splitting step. All vectors have three components on the two-dimensional Cartesian spatial grid (2.5 dimensions or 2D3V). The code uses the finite-difference method to solve for local electric and magnetic fields in space and time.

\subsection{Simulation set-ups}
\begin{center}
\begin{table*}

\caption{\color{black} Overview of the parameters in PIC simulations \footnote{\color{black} In code units, the electron skin depth is $d_e=1$, Debye length, $\lambda_{\rm D}=\sqrt{2\theta_e}d_e$, plasma frequency, $\omega_{pe}=1$, $m_i/m_e=1836$, and the particle-per-cell number is $N_{\rm ppc}=2500$. The temperature gradient scale, $L_{\rm T}$ (more generally, $L_{\rm T}\cos\alpha$), and the notation $\parallel$ is with respect to the direction of the guide field ${B}_0\boldsymbol{\hat{x}}$. The timestep of all the simulations is $0.005~\omega^{-1}_{pe}$ and all boxes are periodic with constant particle number density.}}
\begin{tabular}{|c c c c c c c c c c |} 
 \hline
 $\beta_e$ &$\theta_e$ &$\rho_e=\sqrt{\beta_e}d_e$&$N=N_{\parallel}=N_{\perp}$&$L=L_{\parallel}=L_{\perp}$&$B_0=\sqrt{\frac{2\theta_e}{\beta_e}}$&$\alpha$&$L_{\rm T}$ &$\omega_{ce}=\omega_{pe}\sqrt{\frac{2\theta_e}{\beta_e}}$& $\tau_{\rm comp}$ \\ 
 \hline\hline
  $60$& $0.3$ & $7.75d_e$  & $400$  &$160d_e=20.6\rho_e$ & $0.1$ &$0$& $5000d_e=645.5\rho_e$&$0.1\omega_{pe}$&- \\ 
 \hline
 $60$& $0.3$& $7.75d_e$&$400$ &$160d_e=20.6\rho_e$ &$0.1$& $\frac{\pi}{4}$&$3500d_e=320\rho_e$&$0.1\omega_{pe}$&-\\
 \hline
 $60$& $0.11$& $7.75d_e$ & $400$& $160d_e=20.6\rho_e$&$0.06$&$0$ &$5000d_e=645.5\rho_e$ &$0.06\omega_{pe}$&-\\
 \hline
  $40$& $0.3$ & $6.32d_e$ &$400$ & $160d_e=25.3\rho_e$&$0.1225$ &$0$&$4082.5d_e=645.5\rho_e$  &$0.1225\omega_{pe}$&-\\
 \hline
 $40$& $0.3$ & $6.32d_e$ &$560$ & $225d_e=35.6\rho_e$& $0.1225$&$0$ &$4082.5d_e=645.5\rho_e$ &$0.1225\omega_{pe}$&-\\
 \hline
  $20$& $0.3$& $4.47d_e$ & $400$& $160d_e=35.8\rho_e$&$0.1732$ & $0$&$2886.7d_e=645.5\rho_e$&$0.1732\omega_{pe}$&-\\
 \hline
 $10$& $0.3$ & $3.16d_e$ & $400$ & $160d_e=50.6\rho_e$& $0.245$&$0$ &-&$0.245\omega_{pe}$ &$1007.6\omega_{pe}^{-1}=247 \omega_{ce}^{-1}$\\
 \hline
 $20$& $0.3$ & $4.47d_e$ & $400$&$160d_e=35.8\rho_e$ &$0.1732$ &$0$ &-&$0.1732\omega_{pe}$ &$1425\omega_{pe}^{-1}=247 \omega_{ce}^{-1}$\\
 \hline
 $25$& $0.3$ & $5.0d_e$ & $400$ & $160d_e=32\rho_e$&$0.155$ &$0$ & -&$0.155\omega_{pe}$ &$6455\omega_{pe}^{-1}=1000 \omega_{ce}^{-1}$\\
 \hline
 $20$& $0.3$ & $4.47d_e$ &  $400$& $160d_e=35.8\rho_e$&$0.1732$ &$0$ &$2886.7d_e=645.5\rho_e$ & $0.1732\omega_{pe}$&$4275\omega_{pe}^{-1}=740 \omega_{ce}^{-1}$\\
 \hline
 $20$& $0.11$ & $4.47d_e$ &  $400$& $160d_e=35.8\rho_e$&$0.105$ &$0$& -& $0.105\omega_{pe}$&$2353.2\omega_{pe}^{-1}=247 \omega_{ce}^{-1}$\\
 \hline
\end{tabular}
\label{tab:table}
\end{table*}
\end{center}

We model an electron-proton plasma for a range of plasma $\beta_e$ in a 2.5D box with periodic boundary conditions for particles and electromagnetic waves. The number of particles per cell for each species is $N_{\rm ppc}=2500$, and the density is uniform. The ratio of the electron Larmor radius to the Debye length is $\rho_{e}/\lambda_{\rm D} = \sqrt{\beta_e/2\theta_e} \approx 6\mbox{-}10$, where $\theta_e$ denotes the relativistic temperature of species $s$, defined as $k_{\rm B} T_{s}/m_{s} c^2$, with thermal Lorentz factor $\gamma_{{\rm th}e}\approx \sqrt{1+2\theta_e}=1.26$ (which implies $v_{{\rm th}e}/c=\sqrt{1-1/\gamma^2_{{\rm th}e}}= 0.6$). We use a $400\times400$ box, with each side spanning $(20\mbox{-}36)\rho_{e}$. The magnetic field is along $\boldsymbol{\hat{x}}$, and the temperature-gradient driven force is aligned with it (except in one case). The {\color{black} bulk-velocity gradient} force is associated with a bulk velocity $\boldsymbol{V}_{s}$ in the plane perpendicular to the field. {\color{black} Table \ref{tab:table} summarises all the simulations.}

\subsection{Temperature-gradient force and heat-flux driven whistlers}
\label{sec:hwi}
The heat-flux-driven whistler instability has been discussed in the context of high-$\beta_e$, weakly collisional plasmas such as the intracluster medium and the solar wind. It is well known that when a sufficiently large number of free-streaming, energy-carrying electrons travel down the temperature gradient, parallel whistler waves at the electron Larmor scale become unstable \cite{levinson1992, pistinner1998}. Parallel whistlers do not interact significantly with fast parallel electrons, since in the electron's frame the electric field of the wave rotates opposite to electrons' gyration direction. However, it has now been demonstrated in PIC simulations that the marginal anisotropy (resulting from scattering between parallel whistlers and anti-parallel electrons) can generate oblique or off-axis whistlers \cite{pistinner1998} and scatter electrons in pitch angle and isotropize them in the wave frame \cite{2018JPP_komarov, PhysRevLett_roberg-clark}. Free energy from the gradient is extracted by whistlers and expended in scattering. While oblique whistlers are usually attributed to the marginal electron distribution function of the electrons produced by parallel whistlers, linear theory shows that their growth rate  is not much smaller than that of parallel modes (section 3.3.2 \cite{2024JPP_bott}).
\begin{figure*}
    \includegraphics[width=18cm]{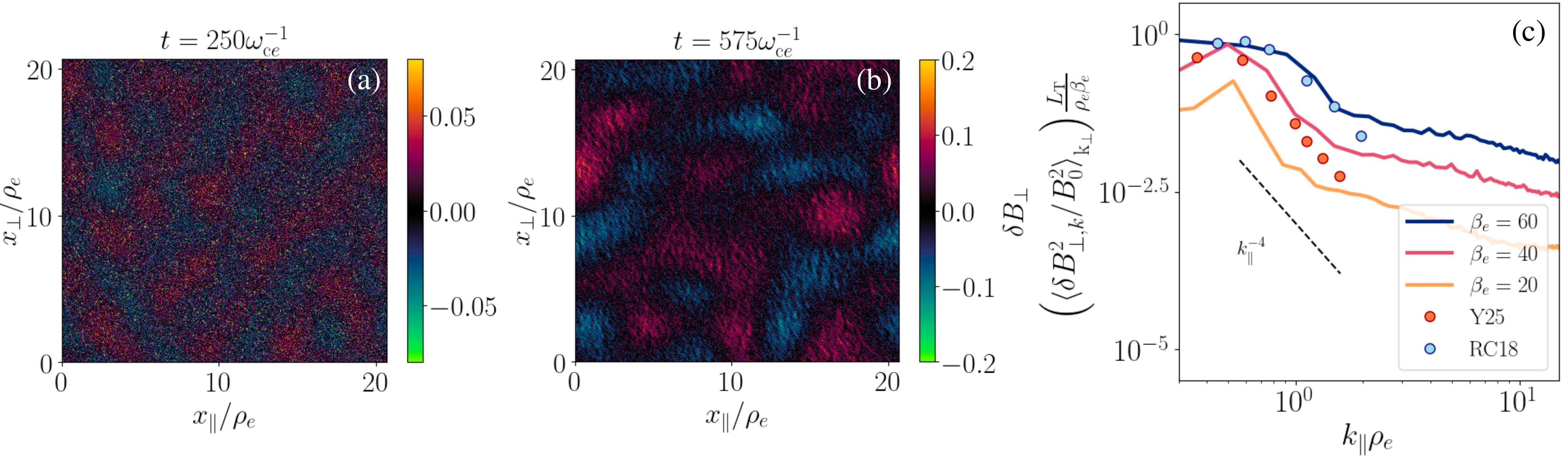}
    \centering
    \caption{Perpendicular out-of-plane component of magnetic field (a) at the onset of whistlers and (b) at saturation for $\beta_e=60$. (c) The spectra of net perpendicular field at saturated stage is shown along with the spectra from previous works (red and blue points {\color{black} from Y25 \cite{yerger2024collisionless} and RC18 \cite{PhysRevLett_roberg-clark} respectively}) for all three simulations. {\color{black} The whistler spectra is peaked at scales $k \sim \rho_e^{-1}$, and at sub-electron scales is consistent with the $k^{-4}_{\parallel}$ power law (dashed black line) observed in prior research. Here, the notation ${\parallel}$ implies relative to the imposed constant magnetic field or $\boldsymbol{\hat{x}}$}.}
    \label{fig:field}
\end{figure*}
\begin{figure*}
    \includegraphics[width=18 cm]{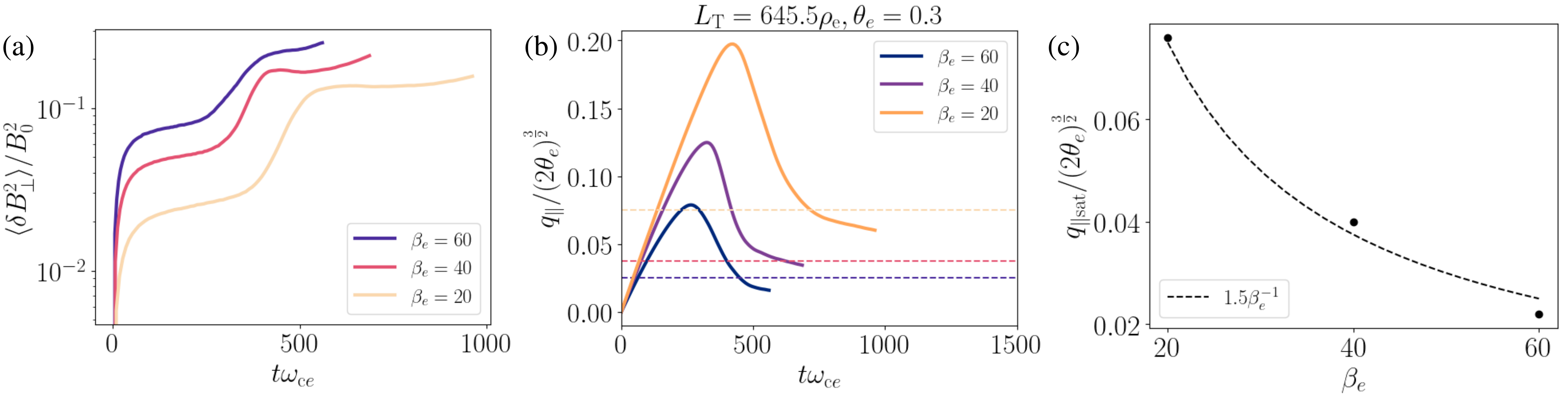}
    \centering
    \caption{The time evolution of (a) box averaged net perpendicular field and (b) parallel heat flux for the three simulations with $\beta_e\in[20,40,60]$. (c) The fitted curve to the saturated parallel heat flux ($1.5\beta^{-1}_e$) with initial $\beta_e$ is shown.}
    \label{fig:time_evolve}
\end{figure*}

We generate the same type of anisotropy using our temperature-gradient force. Fig.~1 in \cite{2024arXiv_yerger} shows the relevant growth rate and frequencies, demonstrating that whistler growth occurs at scales of order ~${\sim}\rho_{e}$ and primarily implies a resonant interaction between thermal electrons and waves. In what follows, we describe two types of simulations: those with aligned and those with misaligned guide fields and temperature gradients.

\subsubsection{Aligned temperature-gradient and magnetic field}
\label{sec:alignedgrad}
Traditional kinetic simulations exploring whistler-regulated heat flux have employed boundary conditions on the two sides of the box along the parallel direction such that a hot (cold) electron population is perpetually maintained. In some simulations, a linearly declining temperature profile is set up across the box at the initial time, while in others the system is evolved with two half-Maxwellians until a steady-state temperature gradient develops. The initial distribution function enables the free streaming of particles from the hot to the cold reservoir. This method is not flexible with respect to general gradient orientation (as is likely in complex astrophysical media) and overheats (overcools) the edges over sufficiently long evolution times.  While our method avoids these issues present in the previous approaches, we first explore the aligned-gradient case in this section. 

Fig.~\ref{fig:wanisotropy} shows a comparison between the analytical expectation ({\color{black} shown below}) and that derived from the PIC box for the parallel momentum-space anisotropy. At short timescales, the Lorentz force term is negligible, and the temperature-gradient will start to drive the distribution $f$ of electrons away from equilibrium in the parallel direction. Over these timescales, the time evolution of $f$ is given by
\ba
\label{eq:Sp1}
\frac{\partial f}{\partial t} =  \frac{p_{\parallel}}{L_{\rm T} \gamma_{\rm p}} \left( \frac{\gamma_{\rm p} -1}{\theta_e} -\frac{5}{2} \right)f_{{\rm M}e}=S_{\rm p} , 
\ea
where $L_{\rm T}=645.5\rho_e$ in all cases with $\beta_e\in[20,40,60]$ considered here for exploring heat-flux-driven whistlers. 
The comparison demonstrates that the parallel anisotropy is correctly reproduced using the temperature-gradient force. The simulations also capture cyclotron resonances, which manifest as hyperbolic dark lines (Appendix \ref{sec:resonance}), as well as particle noise.

Fig.~\ref{fig:field} shows the perpendicular out-of-plane component of the magnetic field at the onset [panel (a)] and at saturation [panel (b)] of the whistler instability. The in-plane and out-of-plane perpendicular components appear similar on visual inspection. Initially, the whistlers propagate along the magnetic field in the plane of simulation. Oblique whistlers emerge eventually, characterized by insignificant propagation and a halt in growth of the parallel heat flux. The parallel-wavenumber spectrum of the net perpendicular magnetic-field fluctuations at saturation [panel(c)] is consistent with spectra from previous simulations (red and blue data points), with its peak at $k \sim \rho_e^{-1}$~\cite{PhysRevLett_roberg-clark,2024arXiv_yerger}. At high $k_{\parallel}$, the spectrum is dominated by noise. We do not find oblique modes beyond an angle of~$\sim\pi/4$ with respect to the parallel direction. 
\begin{figure*}
    \includegraphics[width=18cm]{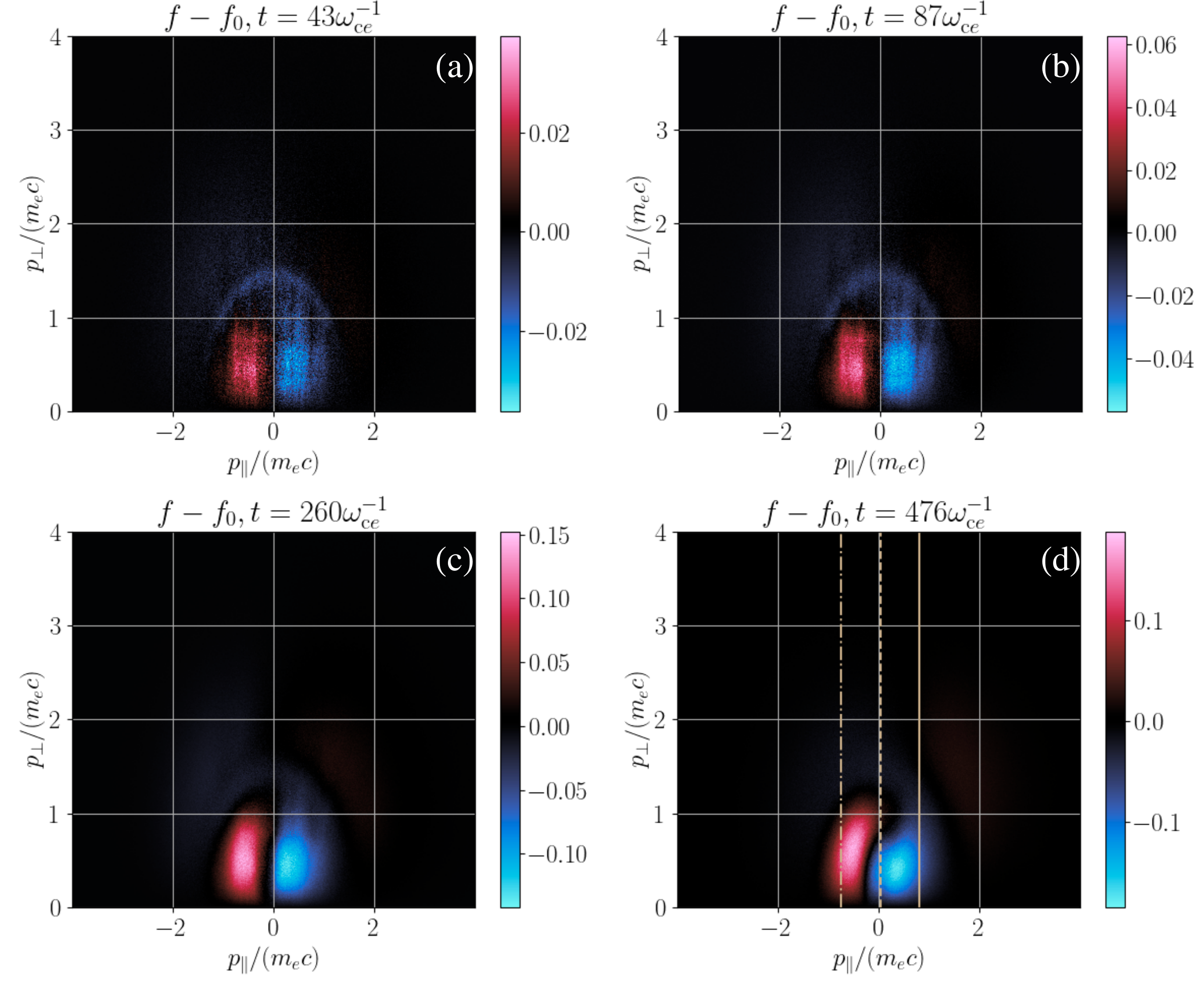}
    \centering
    \caption{The momentum-space anisotropy in the simulation with $\beta_e=20$ at (a) $43\omega^{-1}_{{\rm c}e}$, (b) $87\omega^{-1}_{{\rm c}e}$, (c) $260\omega^{-1}_{{\rm c}e}$, and (d) $476\omega^{-1}_{{\rm c}e}$ (around the time at which the growth of the whistlers ceases).  The three vertical lines in (d) correspond to $\omega/k_{\parallel}=\sqrt{2}\theta_e^{1/2}/\beta_e$ (middle), representing the whistler phase speed with $\theta_e=k_{\rm B}T_{e}/m_{ e}c^2$, and$(\omega+\omega_{{\rm c} e})/k_{\parallel}$ (right) and $(\omega-\omega_{{\rm c}e})/k_{\parallel}$ (left) assume that the characteristic whistler wavenumber satisfies $k_{\parallel} = \rho_{ e}^{-1}$. }
    \label{fig:hwianisotropy}
\end{figure*}

Fig.~\ref{fig:time_evolve} demonstrates the saturated box-averaged perpendicular magnetic-field fluctuations and the parallel heat flux at different plasma $\beta_e$ (see Appendix \ref{app4:theta} for the heat-flux evolution at different values of $\theta_e$). The box-averaged field and the parallel heat flux scale with plasma $\beta_e$, as expected from earlier works. Simulations with larger $\beta_e$ show higher-amplitude waves that suppress the parallel heat flux more efficiently (Appendix \ref{app4:beta} provides additional description of the heat flux). Previous works \cite{PhysRevLett_roberg-clark,2018JPP_komarov} present a similar argument: whistlers scatter electrons in pitch-angle in the wave frame, and heat is therefore advected at the phase speed of whistlers  $v_{\rm ph} = {\omega}/{k_{\parallel}} = {\omega_{{\rm c}e}k \rho^2_{e}}/{\beta_e} \sim {v_{{\rm th}e}}/{\beta_e}$ where the final relation holds for whistlers with $k \rho_{ e} \sim 1$. Thus, $q_{\parallel}\sim m_e n_{ e}v^3_{{\rm th}e}/\beta_e$. Recently, \cite{2024arXiv_yerger} considered the effective collisional rate due to wave-particle scattering that balances whistler linear growth rate, given by $\nu_{\rm eff}\sim \beta_e v_{{\rm th}e}/L_{\rm T}$, and hence $q_{\parallel}\sim n_{ e} v^3_{{\rm th}e} (v_{{\rm th}e}/\nu_{\rm eff})/L_{\rm T}\sim n_{ e}v^3_{{\rm th}e}/\beta_e$. Both effectively consider the resonant wave-particle interaction to be responsible for saturation, predicting a saturated perpendicular field strength $\delta B^2/B^2_0 \sim \nu_{\rm eff}/\omega_{{\rm c}e} \sim \beta_e \rho_e/L_{\rm T}$. This scaling of the perpendicular field amplitude matches within a factor of $2$ for $\beta_e\in[20,40,60]$. 

Fig.~\ref{fig:hwianisotropy} shows the evolution of momentum-space anisotropy at different times for the $\beta_e=20$ simulation. At earlier times, the anisotropy shows signatures of noise (a granulated structure), the nearly vertical dark wave-particle resonant lines, and a resonance feature in the form of a semi-circular arc at $p_{\perp}/m_ec \gtrsim 1$ (\ref{eq:reso}). The anisotropy grows larger until whistlers scatter the particles, such that at saturation it is regulated, as shown between the vertical resonance lines in (d). These indicate the parallel phase speeds corresponding to cyclotron resonance at $k_{\parallel} = \rho_{ e}^{-1}$.   
\begin{figure*}
    \includegraphics[width=18cm]{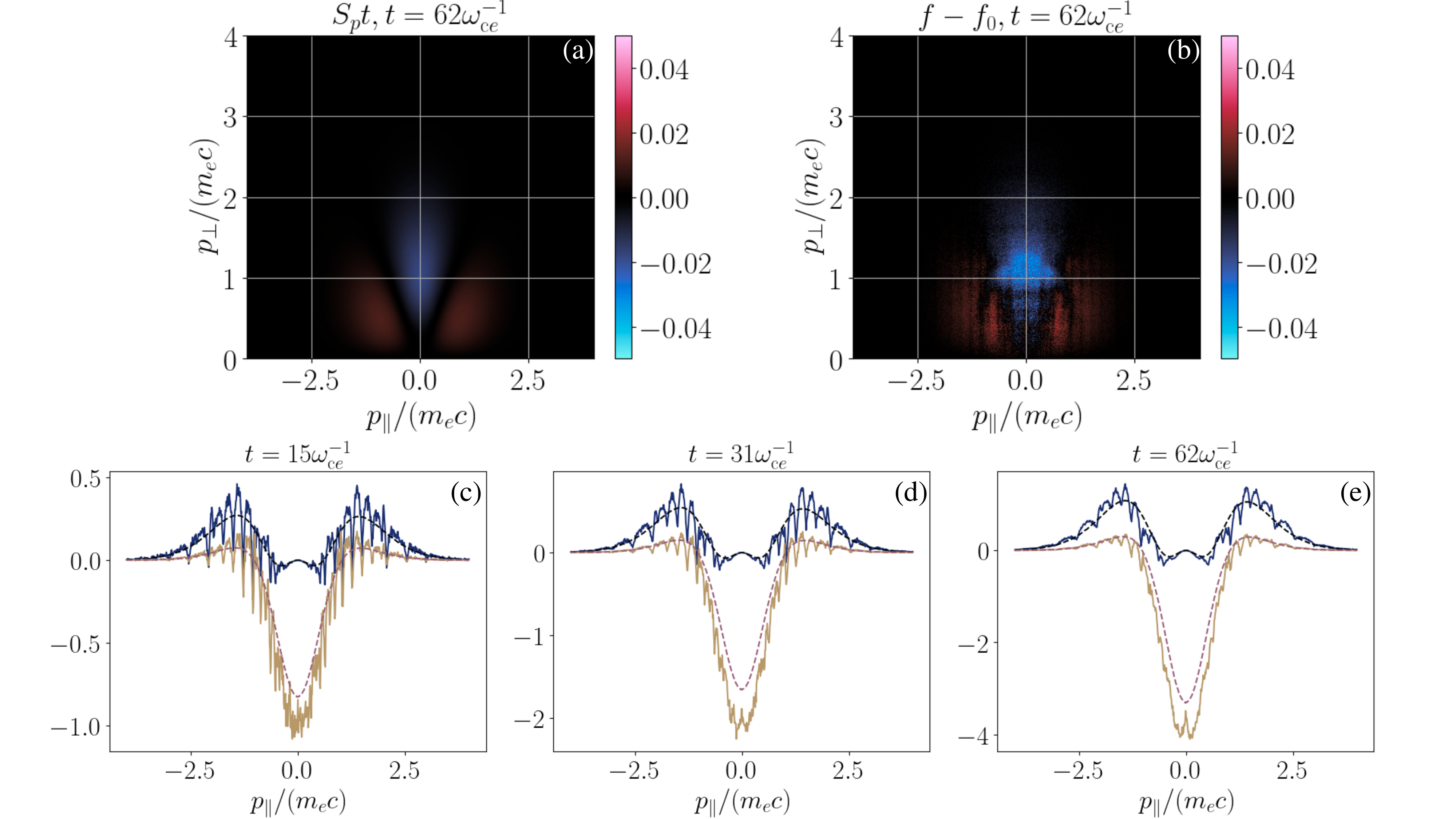}
    \centering
    \caption{Momentum space anisotropy using TF ({\color{black} bulk-velocity gradient}) (a) in our analytical prediction, and (b) in the PIC simulation for a $\beta_e=25$ electron-proton plasma (described by (\ref{eq:Sp2}) in section \ref{sec:est}). In the lower column, we show the following quantities  $p^2_{\parallel}\gamma^{-1}_{\rm p} (f-f_0)$ (solid blue) and $p^2_{\perp}\gamma^{-1}_{\rm p} (f-f_0)$ (solid yellow), $p^2_{\parallel}\gamma^{-1}_{\rm p} S_{\rm p} t$ (dashed blue) and $p^2_{\perp}\gamma^{-1}_{\rm p} S_{\rm p} t$ (dashed yellow), all summed along $p_{\perp}$ direction, at (c) $15\omega^{-1}_{{\rm c}e}$, (d) $31\omega^{-1}_{{\rm c}e}$, and (e) $62\omega^{-1}_{{\rm c}e}$.}
    \label{fig:presanisotropy}
\end{figure*}

\subsubsection{Misaligned temperature-gradient and magnetic field}
\label{sec:misaligned}
In kinetic simulations of whistler-regulated heat flux, misaligned guide magnetic fields and temperature gradients have typically not been studied. In the conventional set-up with maintained thermal-bath boundary conditions, it is non-trivial to maintain a misaligned condition and define meaningful spatial averages. Our method is ideally suited for studying such configurations. This is important to explore, particularly in the presence of both wave-particle scattering and Coulomb collisions, if the theoretical prediction of anisotropic transport (with respect to the local magnetic field) remains. Coherent global magnetic fields are often discussed in astrophysical plasmas, for example in the vicinity of black hole jets or the tentative expectation of coherent fields along cold fronts in clusters. Here, Coulomb collisions are beyond the scope of this paper. But we demonstrate that a misaligned temperature gradient and magnetic field may result in a non-zero diamagnetic heat-flux (see Appendix \ref{app5}). Broadly, this simulation still validates the currently accepted anisotropic heat-flux assumption in the presence of energetically weak but dynamically strong field with $\rho_{ e} < L_{\rm T}$. We use $\rho_e\beta_e/L_{\rm T}\sim 0.1$ and find an increase in one of the perpendicular heat flux components by orders of magnitude (but below the noise level). A larger $\rho_e\beta_e/L_{\rm T}$ (and thus larger amplitudes of whistlers at saturation) is important to understand if the enhancement of perpendicular heat flux is a robust effect. Further the regime $\rho_{ e}<\lambda_{\rm mfp}<L_{\rm T}$ can be explored using TF in the future.

In order to carry out the simulation, we enforce the following temperature-gradient force direction such that $\alpha=\pi/4$ and the force is rescaled using $L_{\rm T}=451.85~\rho_{ e}$ to maintain the same driving of initial heat flux along the field at $\beta_e=60$:
\ba
\boldsymbol{\hat{a}} = {\rm cos} (\alpha)\boldsymbol{\hat{x}} + {\rm sin} (\alpha)\boldsymbol{\hat{y}}
\ea 
We find that the parallel heat flux grows and saturates with the onset of whistlers as expected (Fig.~\ref{fig:misaligned}), almost identical to the case with a pure parallel driving of heat flux at $\beta_e=60$. A plot is added in Appendix \ref{app5} that shows parallel and perpendicular heat fluxes in (a), and the spectra of magnetic field fluctuations in (b). The latter implies a higher power in parallel field fluctuations in the simulation with misaligned TF.

\subsection{{\color{black} Bulk-velocity gradient} force and electron firehose instability}
\label{sec:est}

\begin{figure*}
    \includegraphics[width=18cm]{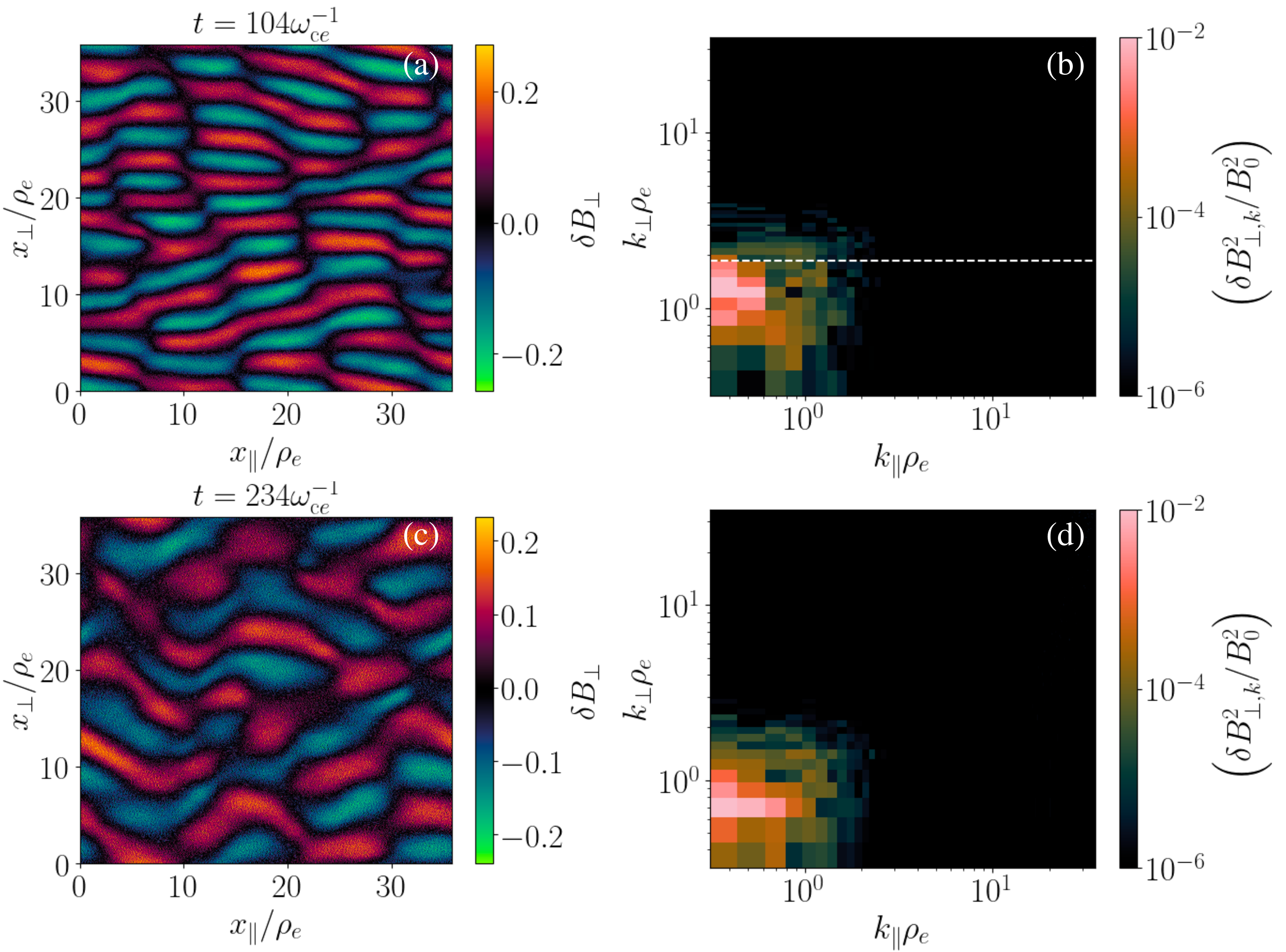}
    \centering
    \caption{The out-of-plane perpendicular magnetic field component at onset of regulation (a) and at saturation (c) for the electron-scale firehose driven by negative pressure anisotropy $\propto 1- 2t/{\color{black} \tau_{\rm comp}}$, where ${\color{black} \tau_{\rm comp}}=246.8~\omega^{-1}_{{\rm c}e}$ and $\beta_e=20$. In the right column, the corresponding 2D spectra of the perpendicular magnetic field fluctuations are shown in (b) and (d) at the same two times. The dashed white lines at onset indicate the scales of peak growth derived analytically [eqs (4.99a) \& (4.99b) in \cite{2024JPP_bott}] for the same temperature anisotropy $\Delta_e$ in the limit of asymptotically large $\beta_e$.}
    \label{fig:field2}
\end{figure*}
 In weakly collisional plasmas, a class of velocity-space anisotropy is discussed frequently in the context of growing (decaying) magnetic field or compression (expansion) of the plasma itself: pressure anisotropy ($p_{\perp}/p_{\parallel}\neq 1$) \cite{2000_hellinger, 2016ApJ_riquelme, Kunz_Abel_Klein_Schekochihin_2018}. Due to conservation of the magnetic moment and second adiabatic invariant, and so $D(p_{\perp}/nB)/Dt = 0$ and $D(p_{\|} B^2/n^3)/Dt = 0$ (where $D/Dt$ is the Lagrangian derivative), anisotropy grows as the macroscopic evolves until electromagnetic instabilities are triggered. Here we discuss electron firehose instability, which is triggered when $p_{e\perp}/p_{e\parallel}<1$ (for the remainder of this section, we drop the $e$ subscripts).

In collisionless, homogeneous plasma, two types of electron firehose instability have been discussed in linear theory and PIC simulations \cite{2000_li, 2003_gary}: a resonant, oblique instability, and a parallel propagating non-resonant instability (although \cite{2008JGRA_hellinger} claimed cyclotron resonance plays a role in the latter). 
The long-wavelength firehose can be conceptualized as the destabilization of Alf\'ven waves by the reduction of magnetic tension by pressure anisotropy (the fluid-firehose instability \cite{1958RSPSA_chandra}), while the oblique firehose can be  described as an instability of kinetic Alf\'ven waves (KAW; a short-wavelength extension of the fluid version). Although there is debate regarding the role of wave-particle resonance, most previous works agree that the electron distribution is driven to marginal stability by the back-reaction of the instability: $\Delta_{ e} = p_{\perp}/p_{\parallel} \simeq 1 \sim -\mathcal{A}/\beta_e$, for $\mathcal{A}$ an order-unity constant. 

\begin{figure*}
    \includegraphics[width=18cm]{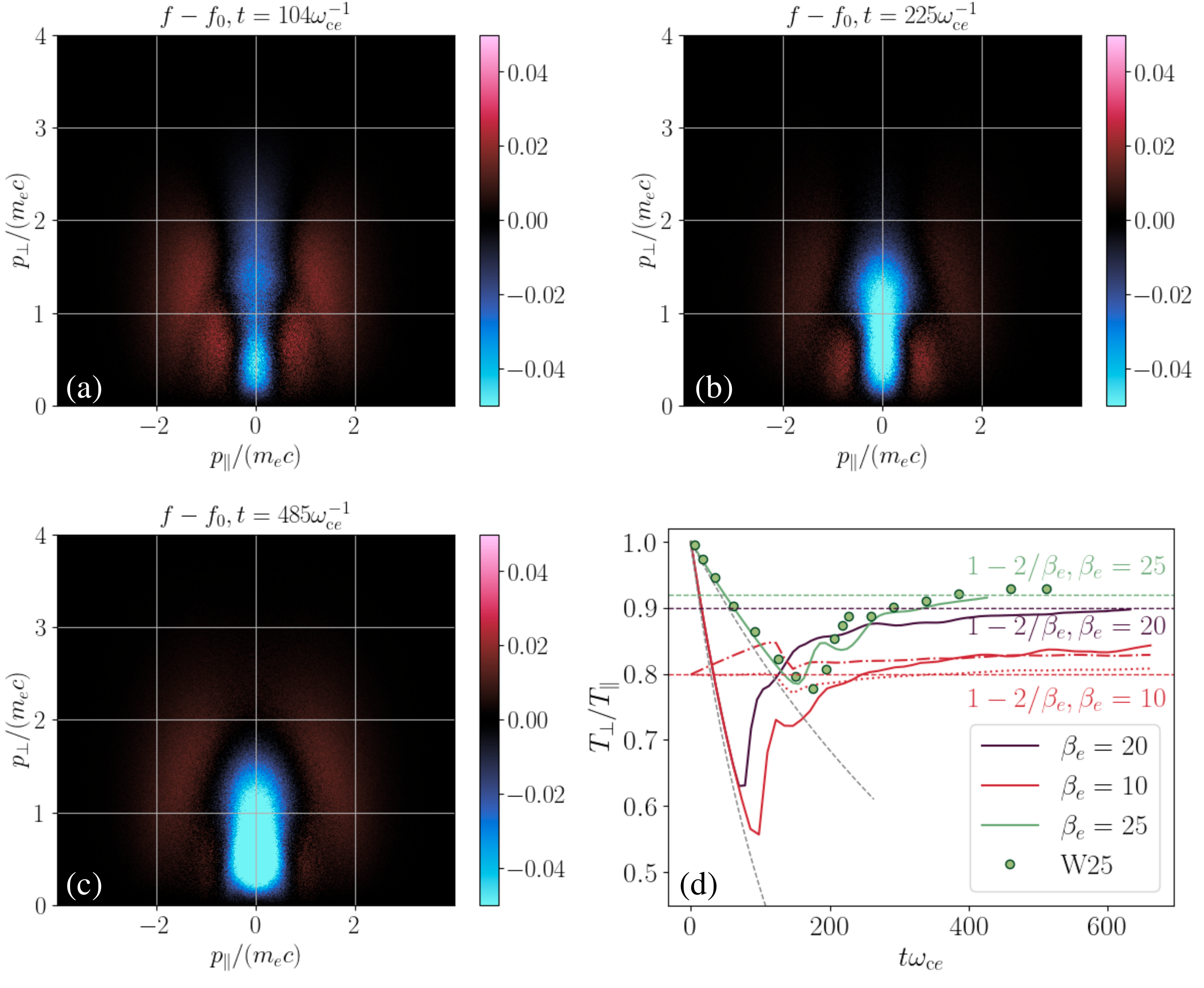}
    \centering
    \caption{The momentum space anisotropy at the (a) onset of regulation, (b) around the time of saturation, and (c) post-saturation with $\beta_e=20$ and ${\color{black} \tau_{\rm comp}}=246.8~\omega^{-1}_{{\rm c}e}$. (d) The relative perpendicular to parallel temperature from the same simulation is shown in solid maroon line with comparisons of $\beta_e=10$ (solid red line), $\beta_e=25$ and ${\color{black} \tau_{\rm comp}}=1000~\omega^{-1}_{{\rm c}e}$(solid green line), and a non-relativistic simulation from a work in preparation, {\color{black} W25} \cite{Winarto} (green circles) with the same parameters as the last. All the simulations saturate at about $1-2/\beta_e$ defined by initial $\beta_e$, except for the lowest $\beta_e$ simulation in which the marginal temperature anisotropy follows $1-2/\beta_{e,\parallel}$ defined dynamically by the red dot-dashed line. The red dotted line shows dynamical $1-2/\beta_{e}$ for $\beta_e=10$ simulation.}
    \label{fig:paniso}
\end{figure*}

Our setup includes a {\color{black} bulk-velocity-gradient}-driven force with ${\color{black} \tau_{\rm comp}}\in[246.8, 740.4, 1000]~\omega^{-1}_{{\rm c}e}$, for $\beta_e$ between $10$ and $25$, associated with a bulk velocity $\boldsymbol{V}_{s}$ (\ref{BV_field_perp}) in the perpendicular plane. In this section, we exclude the temperature-gradient-driven force and verify the characteristics of the oblique firehose instability. Fig.~\ref{fig:presanisotropy} shows the anisotropy driven for ($\beta_e=25$ and ${\color{black} \tau_{\rm comp}}=1000~\omega^{-1}_{{\rm c}e}$) over short timescales and a comparison with the analytical evolution in the absence of the Lorentz force (at early times). The change in the distribution function due to the ({\color{black} bulk-velocity gradient}) TF is
\ba
\label{eq:Sp2}
\frac{
\partial f}{\partial t} = \left(\frac{4}{3}p^2_{\parallel} - \frac{2}{3} p^2_{\perp} \right)\frac{f_{{\rm M}e}}{2\theta_e {\color{black} \tau_{\rm comp}} \gamma_{\rm p}}= S_{\rm p} . \label{VGTF_earlytimes}
\ea
The upper panels show the anisotropy in two-dimensional momentum space as predicted by (\ref{VGTF_earlytimes}) [panel (a)], and what we find in the PIC simulation [panel (b)]. In the simulation, many nearly vertical resonant lines and a dominant semi-circular resonance arc at $p_{\perp}/m_ec \gtrsim 1$ are visible (see Appendix \ref{sec:resonance}), similar to those seen in simulations with the temperature-gradient-driven TF. The lower panels show the one-dimensional profiles of the phase-space anisotropy, weighted by the parallel and perpendicular energy carried by the electrons and integrated in the perpendicular momentum, at three different times. The good agreement verifies that the anisotropy is driven as expected over short timescales, although there is a small deviation from the analytical prediction at small momenta ($\left| p/m_ec \right|\lesssim 1$) in both the parallel and perpendicular integrals,  due to the circular resonant feature, which produces an additional deficit of electrons within that momentum range. 

Fig.~\ref{fig:field2} shows the out-of-plane component of the perpendicular magnetic field (which is dominant in amplitude) at (a) the onset of regulation and (c) near saturation of the instability. The modes are quasi-perpendicular until saturation. There is no significant temporal propagation of these modes with time in our simulations, {\color{black} as predicted by linear theory~\cite{2024JPP_bott}}. The corresponding spectra at these two stages [panels (b) and (d), respectively] indicate the evolution of the modes from small-scale, quasi-perpendicular structure ($k_{\parallel}\rho_e < 1 \lesssim k_{\perp}\rho_e$) to reduced obliqueness at larger scales ($k_{\parallel}\rho_e \approx k_{\perp}\rho_e \lesssim 1$). 
The peak pressure anisotropy observed before regulation in this simulation (${\color{black} \tau_{\rm comp}}=246.8~\omega^{-1}_{{\rm c}e}$) is $\Delta_{e}\beta_e = -7.6$. 

Fig.~\ref{fig:paniso} shows the 2D momentum space anisotropy at three different times: (a) the onset of regulation, (b) the onset of saturation, and (c) post-saturation.  The phase-space anisotropy evolves towards a marginal level due to resonant interactions (the black arc), until it is confined to small momenta/energy. The perpendicular-to-parallel electron temperature ratio (related to the temperature anisotropy $\Delta_e$ through $T_{\perp}/T_{\parallel}=\Delta_e+1$) is shown as a function of time for multiple cases in (d). The horizontal dashed lines indicated marginal temperature-anisotropy thresholds proposed by theory and calculated using the initial magnetic field. The temperature anisotropy initially evolves as $\Delta_e \simeq 1-2t/{\color{black} \tau_{\rm comp}}$ (the analytical prediction is derived in Appendix \ref{app6}), indicated by the gray dashed lines. We include simulations with fast plasma expansion (short ${\color{black} \tau_{\rm comp}}$) with initial $\beta_e \in [10,20]$ (red and dark maroon lines). Both saturate at the expected threshold. However, in the simulation with $\beta_e=10$, the threshold is better matched by the dynamic $1-2/\beta_{e,\parallel}$ (thin dot-dashed red line) where $\beta_{e,\parallel}=2 n_{ e} T_{\parallel}/\langle B^2 \rangle$. This is most plausibly due to saturation occurring with a mean magnetic field that is weaker than at the start by a significant fraction of the initial field. We also compare two equivalent simulations: a non-relativistic case by \cite{Winarto} with $\beta_e=25$ and ${\color{black} \tau_{\rm comp}}=1000~\omega^{-1}_{{\rm c}e}$,  and a corresponding relativistic case using our method. The maximum anisotropy is comparable between the two (solid green line and green circles), although we see a mildly faster regulation in our simulation, possibly due to the additional resonant interactions we discuss in Appendix \ref{sec:resonance}. 
\begin{figure}
    \includegraphics[width=8.5cm]{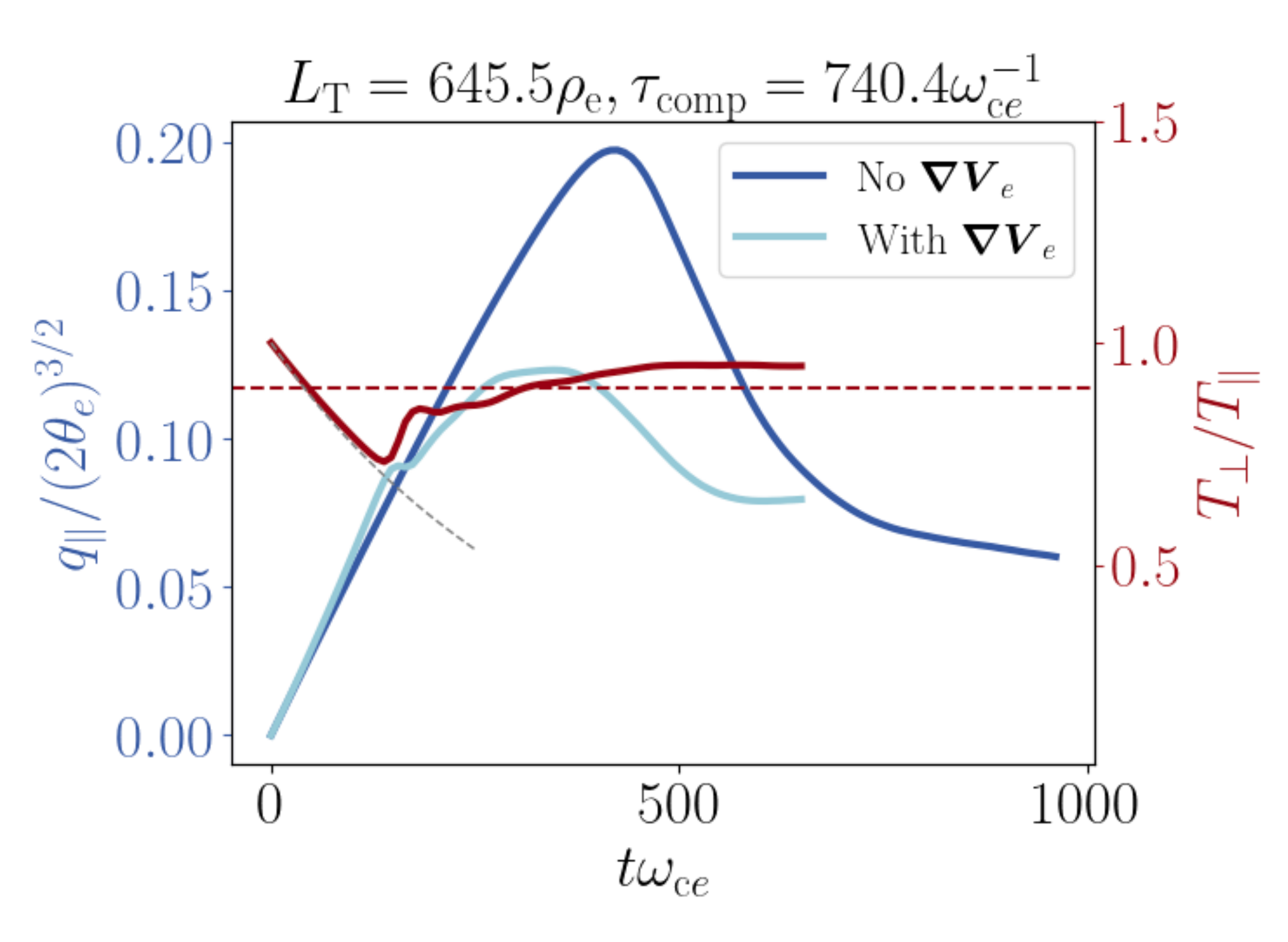}
    \centering
    \caption{{\color{black} Heat flux (blue and cyan) and pressure anisotropy (maroon) are plotted along twin y-axis and normalised time along x-axis for simulations (Table \ref{tab:table}) with only driven parallel heat flux ($q_{\parallel}$) and both driven parallel heat flux and pressure anisotropy ($\Delta_e, q_{\parallel}$).} Heat flux suppression does not depend on the whistlers (cyan line) in this case and oblique firehose plays the key role validated by the simultaneous regulation of pressure anisotropy.}
    \label{fig:collective}
\end{figure}
\subsection{Collective effect of heat-flux driven whistlers and pressure anisotropy driven firehose instabilities}
One of the unresolved issues in the context of transport models and fluid closure is that the marginal momentum-space anisotropy in the presence of multiple kinetic instabilities (driven by both temperature gradients and {\color{black} bulk-velocity gradients}) cannot be quantified with current theory or simulations. Thus, our method provides a key step towards building an accurate fluid-closure model. In addition, the collective evolution of these instabilities opens the possibility of new saturation mechanisms. Here we demonstrate that driving a parallel heat flux and pressure anisotropy simultaneously is enabled by the TF-PIC method, and provide evidence that the presence of more than one kinetic instability interacting with the particles may result in different mechanisms and timescales for the regulation of fluxes. 

To characterize the relative strength of the forcing, or equivalently the timescales over which the two free-energy sources act on the plasma, we introduce the parameter 
\ba
\nonumber 
\mathcal{Q} &\equiv & \frac{L_{\rm T}}{v_{{\rm th}e} {\color{black} \tau_{\rm comp}}} \\
&=& \frac{{{L}_{\rm T}}/{\rho_{ e}}}{{\color{black} \tau_{\rm comp}}\omega_{{\rm c}e}}
\ea
This ratio is thought to be of order of unity (section 2.2.3 \cite{2024JPP_bott}) when temperature fluctuations in the medium are passively advected with velocity fluctuations. However, in realistic plasmas this ratio will span a range depending on the smallest scales attained by temperature fluctuations due to fluid instabilities and radiative processes. In this analysis, we run simulations with the parameters in section \ref{sec:hwi} with $\beta_e=20$ and section \ref{sec:est}, using a ratio $\mathcal{Q}\sim 0.9$ in Fig.~\ref{fig:collective}, such that the free energy sources are of similar strength.  Naively, one might expect no difference in the suppression mechanism of heat flux in this scenario. However, we instead find that the saturated heat flux is due to oblique firehose modes instead of heat-flux driven whistlers, such that the parallel heat flux ($q_{\parallel})$ is suppressed compared to free-streaming heat-flux ($q_{\rm fs}$) in the following way: 
{\color{black}
\begin{equation}
 q_{\parallel}\sim \lambda_e q_{\rm fs}/L_{\rm T} \sim (L_{\rm V}/\beta_e) q_{\rm fs}/L_{\rm T} = q_{\rm fs}/\mathcal{Q}\beta_e  .  
\end{equation} 
}
The dark blue line shows whistler-regulated heat-flux evolution, and the light blue line shows firehose-regulated heat-flux evolution. The final saturated flux is not identical to the case without firehose, and the regulated temperature anisotropy is above the marginal state ($> -2/\beta_e$). Despite similar strengths of the driving by macroscopic sources of phase-space anisotropies, a wider spectrum of firehose instabilities is triggered as the pressure anisotropy evolves from small to large values. This may lead to resonant scattering even before whistlers become unstable. We will explore the physical relevance and astrophysical regimes of this parameter in detail in a forthcoming paper \cite{Choudhury}.

\section{Discussion}
\label{sec:disc}
In this work, we have introduced a new theoretical approach for modeling transport in weakly collisional, magnetized, {\color{black} high-$\beta$} plasmas: thermodynamic forcing. Macroscopic transport of momentum and heat results from anisotropies in the distribution function, which themselves arise due to gradients of macroscopic properties (e.g. temperature or velocity). The thermodynamic forcing method involves introducing an anomalous force that produces anisotropies in homogeneous plasmas, approximating those produced by macroscopic gradients in inhomogeneous weakly collisional plasmas. Transport is ultimately determined by the regulation of this anisotropy through both Coulomb collisionality and anomalous scattering due to kinetic instabilities. We have shown that the approximation's error scales with the total collisionality. {\color{black} In the most general form that we have derived, thermodynamic forcing can be adjusted to mimic the effect on the distribution function of all relevant macroscopic gradients — temperature, velocity, and magnetic-field gradients — that arise during the dynamical evolution of a weakly collisional plasma. Temperature and velocity gradients are each modeled with their own specific forms of thermodynamic forcing, while magnetic-field gradients enter through the  electron-ion-drift and friction-force already contained in the most general formulation. This provides a pathway towards systematic, first-principles transport modeling in such plasmas, including viscosity, heat conductivity, and resistivity. We emphasize that the primary contribution of this work is the formulation, implementation, and validation of the thermodynamic forcing framework itself. The simulations and examples presented here are therefore not intended to exhaustively characterize transport coefficients or instability saturation in new regimes, but rather to establish that the method faithfully reproduces known behavior, as well as being applicable in scenarios that are difficult to study using existing kinetic simulation approaches.} {\color{black} While we propose and validate thermodynamic forcing for Maxwellian plasmas, conceptually, our approach can be generalised to plasmas near a collisionless (non-Maxwellian) equilibrium \cite{ewart2025relaxation}.}

Beyond proposing thermodynamic forcing, we have implemented it numerically for temperature and velocity gradients: first, in test-particle simulations, then in full PIC (`TF-PIC') simulations. Our test-particle simulations confirm that our method does not come with any unexpected numerical instabilities. However, the temperature-gradient-driven thermodynamic force can cause (unphysical) runaway acceleration of certain particles if applied over time intervals much longer than its formal regime of validity. We observe the emergence of electromagnetic instabilities in our TF-PIC simulation with two representative examples from the classes of temperature-gradient-driven and velocity-gradient-driven instabilities, namely, the whistler-heat-flux (section \ref{sec:alignedgrad}) and the electron firehose (section {\ref{sec:est}) instabilities. The results of our TF-PIC simulations are consistent with previous PIC simulations that explicitly include macroscopic gradients, verifying that the method is suitable for studying the same transport phenomena. Furthermore, we have, for the first time, explored transport in plasmas with both temperature and velocity gradients directly using the TF-PIC method, revealing a previously unrecognized interplay between distinct unstable modes. We have also used TF-PIC for the first time to explore temperature gradient misaligned with the background magnetic field and we find evidence of non-zero diamagnetic heat-flux facilitated by whistler instability.

This study has some limitations. We have focused on demonstrating the key features of our method using relativistic thermodynamic forcing in a plasma with a relativistic temperature $\theta_e=0.3 ~(\gamma_{{\rm th}e} = 1.26)$, primarily due to numerical cost. As a result, certain physical processes -- for example, resonant scattering -- are subject to relativistic corrections that may affect some of our results quantitatively, but not qualitatively. We anticipate that applying thermodynamic forcing with a Boris pusher at smaller values of $\theta_e$ -- instead of our more generalized and less expensive Vay-pusher implementation -- will mitigate this issue. Another limitation is that, for the relatively small values of $L_{\rm T} \lesssim 200 \rho_e$ used in these simulations, the temperature-gradient-driven thermodynamic force can drive a runaway heat flux. This issue can likely be mitigated by running with larger simulation domains or by damping the force at suprathermal particle velocities or by weak Coulomb collisions. We plan to explore these solutions in future work. Finally, although we claim that this method can be used to model transport, generally, it is only valid for domains that are much smaller than global scales. Thus, this method is not suitable for simultaneously modeling fluid plasma instabilities, as actual gradients and the effect of the force could combine in unphysical ways. However, such a combined approach is not possible with PIC simulations using available computational resources. This is precisely why we propose incorporating the effect of large-scale gradients into a small periodic domain of plasma with thermodynamic forcing, to accurately capture particle distributions and determine transport. 

The thermodynamic forcing method addresses key challenges in previous studies of transport in weakly collisional plasmas, in which the macroscopic gradients were included directly. First, in such approaches, the size of the simulation domain literally sets the gradient \cite{PhysRevLett_roberg-clark}; so, achieving realistic scale separations requires increasingly expensive simulations, an issue avoided by thermodynamic forcing. Additionally, the method has the flexibility to accommodate complex geometries, and multiple types of gradient simultaneously. Indeed, our initial results suggest that driving multiple kinetic instabilities can significantly change our understanding of which instability is dominant and mediating transport -- a topic which we will explore in detail in \cite{Choudhury}. The thermodynamic forcing method, when combined with Coulomb collisionalilty, is an ideal tool for studying the transition from classical transport models to those that include anomalous scattering between electromagnetic instabilities and electrons/ions. Finally, there is strong potential to leverage machine learning using the data from a large number of TF-PIC simulations -- spanning various macroscopic gradients, collisionality, and magnetizations -- to develop analytical and statistical models of transport that correctly incorporate anomalous scattering physics. Thus, our research represents a significant step towards solving the problem of transport in weakly collisional, magnetized plasmas. 

\section*{Acknowledgements}
The authors acknowledge Alex Schekochihin and Matt Kunz for discussions at the initial stage of conceptualization of this work. PPC thanks Chris Reynolds for many useful discussions on whistler-regulated heat flux and associated collisionality. The authors also acknowledge M. Vranic and P. Bilbao in the {\tt OSIRIS} team for initial input into the code. PPC acknowledges J. Drake, M. Swisdak, and G. Roberg-Clark for several discussions on the PIC method in {\tt P3D} code to study heat flux-driven whistlers. This research was supported primarily by the UKRI (grant number MR/W006723/1) and also in part by grant NSF PHY-2309135 to the Kavli Institute for Theoretical Physics (KITP). All simulations reported in this work have been carried out with {\tt OSIRIS} using high-performance computing resources from Advanced Research Computing (ARC), University of Oxford.
\appendix

\color{black}
\section{Special relativistic coordinate transformation}
\label{app0}
The identity (\ref{eq:rel_v_add}) that relates particles' 3-velocities in the laboratory frame and the frame of the fluid which has a non-relativistic bulk velocity $\boldsymbol{V}_s$ is used to carry out the coordinate transformations in section \ref{sec:covmethod}. To derive it, we note that if $v^\prime_s$ is the speed of a particle as measured from the fluid frame, then the same quantity measured in the laboratory frame is
\ba
v = \frac{v^\prime_s + V_s}{1+ {v^\prime_s V_s}/{c^2}} . 
\ea
Then the Lorentz factor $\gamma_{\rm p}$ for a particle in the laboratory frame is given by
\ba
\nonumber
\gamma^2_{\rm p} &=& \frac{1}{1-{v^2}/{c^2}} \\
\nonumber
&=& \frac{c^4({1+ {v^\prime_s V_s}/{c^2}})^2}{(c^2-v^{\prime2}_s)(c^2 - V^2_s)} . \\
\ea
Rearranging gives
\ba
u &=& \gamma_s u^\prime_s + \gamma_{\rm p} \gamma_s V_s , 
\ea
where $u=\gamma_{\rm p}v$ is the magnitude of the 3-velocity, and we have used $\gamma_{\rm p}\approx\gamma^\prime_{\rm p}$, since the Lorentz factor associated with fluid velocity is $\gamma_s\approx1$. The identity (\ref{eq:rel_v_add}) follows. We note that this modifies the definition of peculiar 3-velocity when the particles are allowed to be relativistic.

\color{black}
\section{The source term associated with {\color{black} bulk-velocity gradient} at high $\gamma_{\rm p}$}
\label{app1}
The {\color{black} bulk-velocity gradient} driven source term has two parts in the relativistic case. The analogue of the non-relativistic case contributes to the shear/compression tensor as discussed earlier. At $\gamma_{\rm p} \gg 1$, we find an additional $\gamma_{\rm p}$ dependent term that is not relevant to the weakly relativistic case. Here we will calculate the force derived from this source term and associated with {\color{black} the divergence of fluid velocity,} $\boldsymbol{\nabla}\cdot\boldsymbol{V}_{s}$. The differential equation to be solved for this force, $F_1 \hat{\boldsymbol{v}}$, is 
\ba
\nonumber
F^{\prime}_1 - \frac{F_1}{\theta_s} = \frac{m_{s}c \boldsymbol{\nabla}\cdot \boldsymbol{V}_{s}}{\theta_s} \left(-\frac{2}{3}\frac{\gamma_{\rm p} - 1}{\sqrt{1 - \gamma^{-2}_{\rm p}}} + \frac{1}{3} \frac{\gamma_{\rm p} - \gamma^{-1}_{\rm p}}{\sqrt{1 - \gamma^{-2}_{\rm p}}}\right), 
\ea
where the prime denotes the derivative with respect to the Lorentz factor. This can be solved by using an integrating factor $e^{-(\gamma_{\rm p}-1)/\theta_s}$ and changing to a new variable $w=\gamma_{\rm p}-1$ as follows:
\ba
\nonumber
F_1 &=& e^\frac{{(\gamma_{\rm p }-1)}}{\theta_s} \frac{m_{s}c \boldsymbol{\nabla}\cdot \boldsymbol{V}_{s}}{\theta_s}\int^{\infty}_{\gamma_{\rm p}-1} \mathrm{d}w \; e^{-\frac{w}{\theta_s}}\left(\frac{2}{3}\frac{w(w+1)}{\sqrt{w(w+2)}} \right.\\
&&\left. \qquad \qquad \qquad \qquad -\frac{1}{3} \sqrt{w^2 + 2w} \right) .
\ea

Each term in the RHS can be integrated for two regimes $\gamma_{\rm p} - 1 > \theta_s$ or $\gamma_{\rm p} - 1 \leq \theta_s$, the latter is non-relativistic. In that limit, the integral goes to zero. In the former limit, the dominant contribution is in the vicinity of $\gamma_{\rm p} -1$ due to exponential cut-off.
\ba
\nonumber
F_1 &=& e^{\frac{(\gamma_{\rm p }-1)}{\theta_s}} \frac{m_{s}c \boldsymbol{\nabla}\cdot \boldsymbol{V}_{s}}{\theta_s}\int^{\infty}_{0} \mathrm{d} \epsilon \; e^{-\frac{(\gamma_{\rm p} -1 +\epsilon)}{\theta_s}}\\
\nonumber
&&  \times \left (\frac{1}{3}\frac{\sqrt{\gamma_{\rm p} -1 +\epsilon} (\gamma_{\rm p} - 1+\epsilon)}{\sqrt{\gamma_{\rm p} +1 +\epsilon}} \right) \\
\nonumber 
\\
&=& m_{s} \frac{\boldsymbol{\nabla}\cdot{\boldsymbol{V}_{s}}}{3} (\gamma_{\rm p} \boldsymbol{v}) \left(\frac{\gamma_{\rm p}-1}{\gamma_{\rm p} +1} \right).
\ea

\begin{figure*}
    \includegraphics[width=17cm]{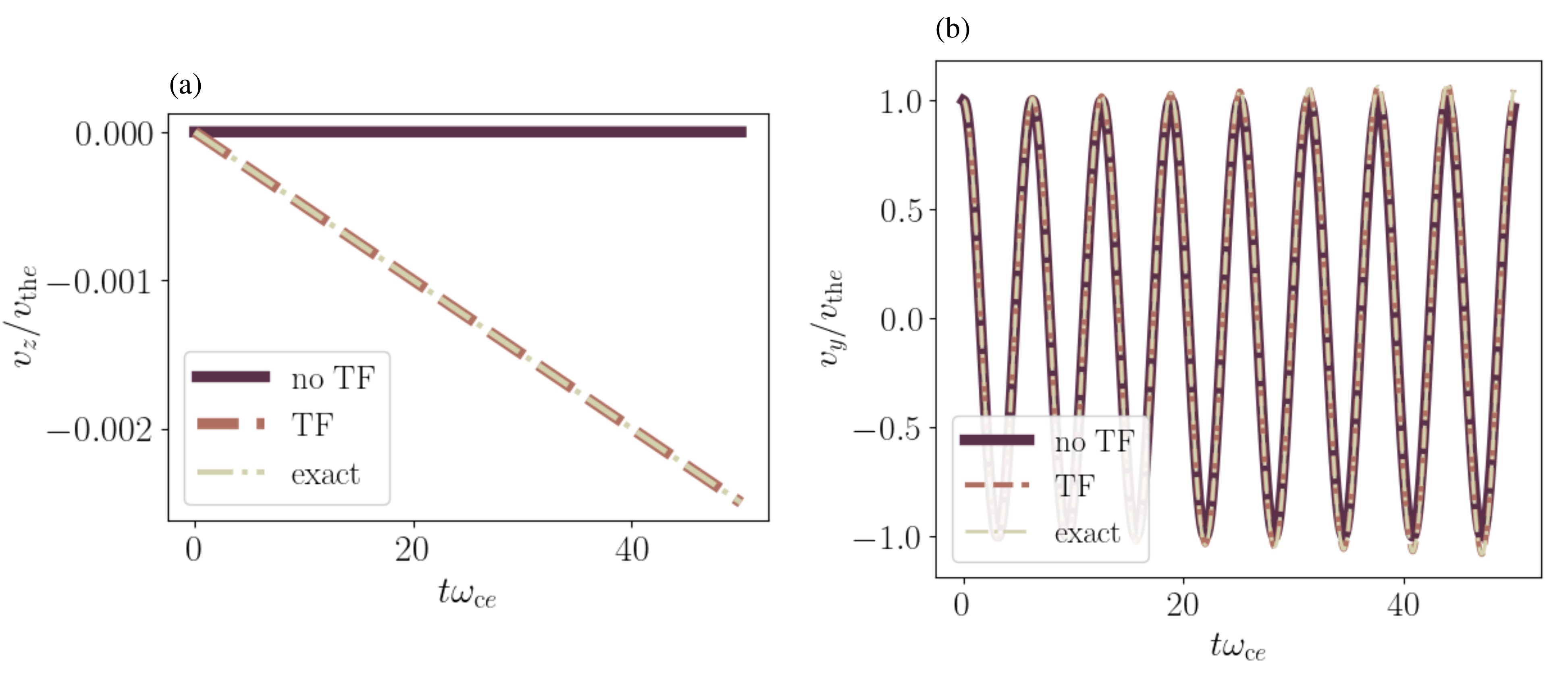}
    \centering
    \caption{Single particle Boris pusher modified to use TF. The left and right panels show effect of thermal gradient and {\color{black} bulk-velocity gradient} in ${v}_{\rm parallel}$ and ${v}_{\rm y}$ (one of the perpendicular components) respectively. The parameters used are the following: thermal gradient scale ${L}_{\rm T}=5000 \rho_{ e}$, ${\color{black} \tau_{\rm comp}}= 200 \omega^{-1}_{{\rm c}e}$. The left panel corresponds to a ${v}^2_{\rm r}={v}^2_{\perp} - \frac{3}{2}v^2_{{\rm th}e}<0$ with ${v}_{\rm y} = v_{{\rm th}e}$.}
    \label{fig:boris_velo}
\end{figure*}
\begin{figure}
    \includegraphics[width=8.5cm]{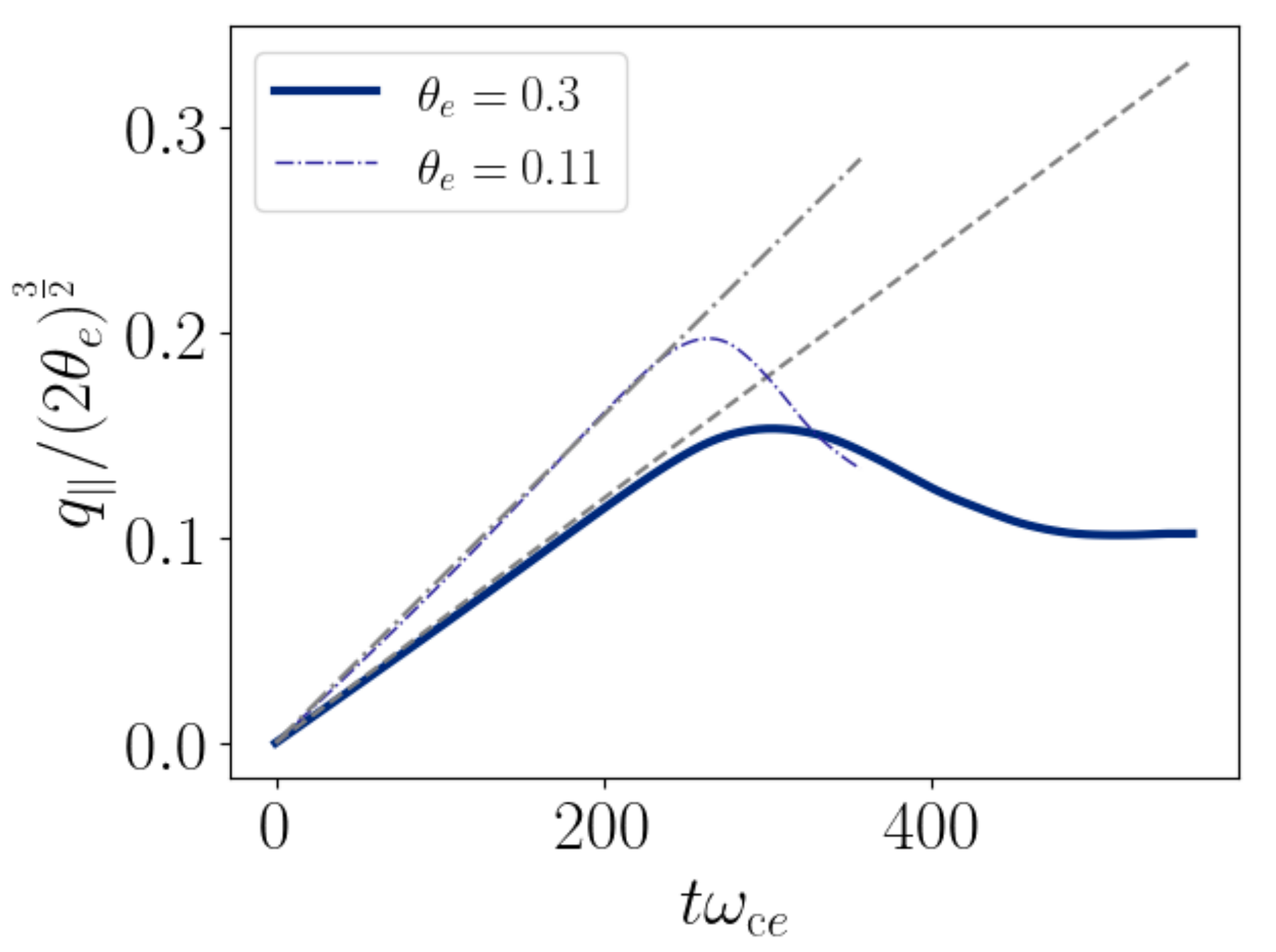}
    \centering
    \caption{Heat flux evolution with time at same $\beta_e$ and different $\theta_e=k_{\rm B}T_e/m_ec^2$. The gray lines denote the approximate $\theta_e$ dependent slopes calculated in Appendix \ref{app4:theta}.}
    \label{fig:htflx_theta}
\end{figure}
\begin{figure*}
    \includegraphics[width=17cm]{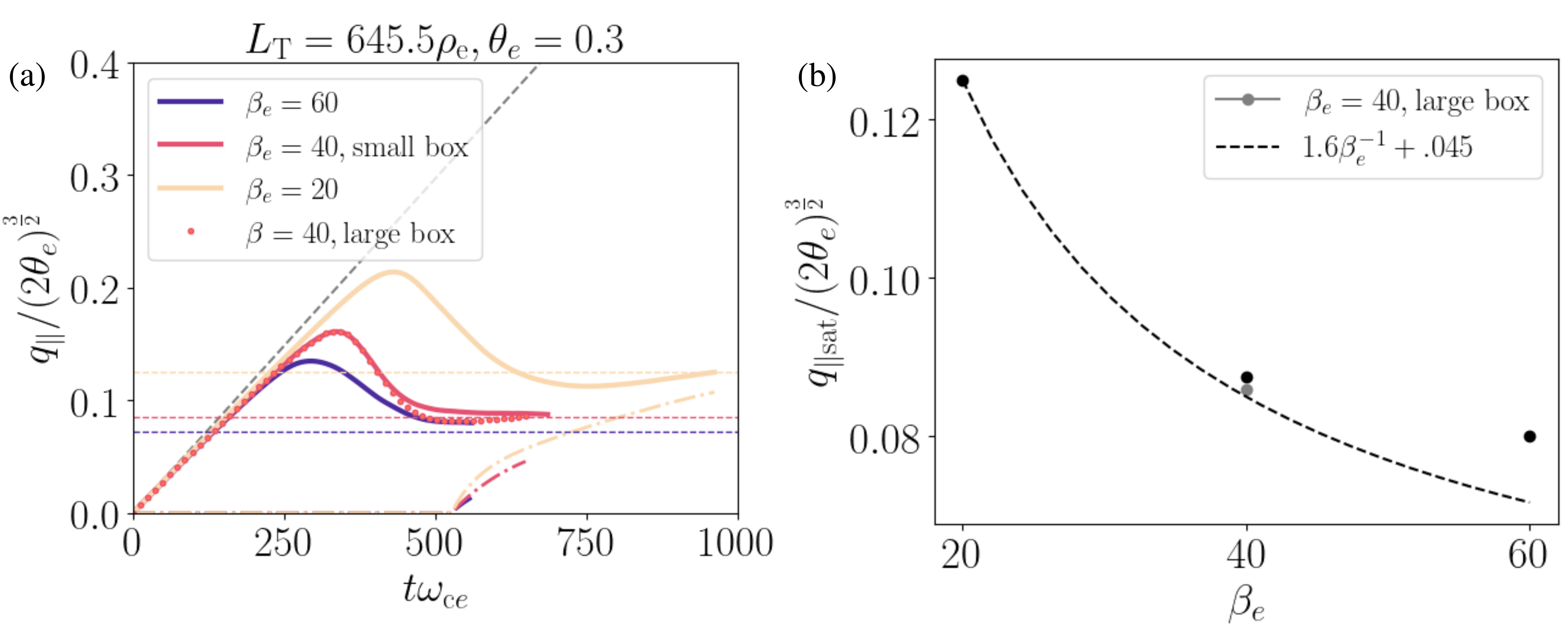}
    \centering
    \caption{(a) The net parallel heat flux across the entire momenta space for three simulations with $\beta_e \in [20,40,60]$ in solid lines, and net contribution to it from runaway electrons (dot-dashed lines). The gray line shows the analytical growth of parallel heat flux due to TF and dashed horizontal lines represent $1.6\beta^{-1}_e$ scaling. (b) The fitted curve to the saturated net parallel heat flux.}
    \label{fig:htflx_all}
\end{figure*}
\begin{figure}
    \includegraphics[width=8.cm]{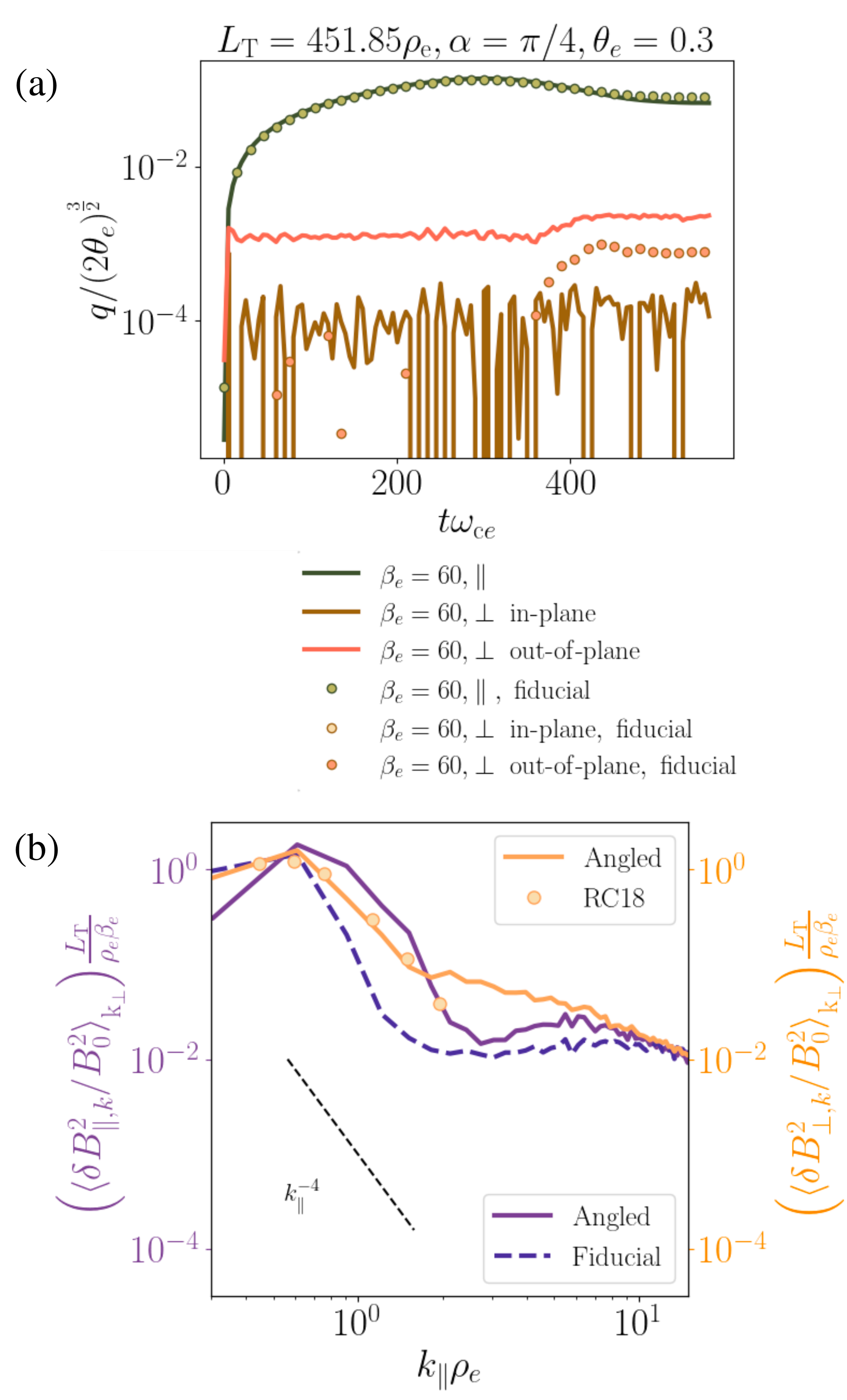}
    \centering
    \caption{(a) Heat flux in parallel and perpendicular directions to the magnetic field ($\beta_e=60$) for the simulation with force driving equal heat flux along and across the magnetic field. The growth of whistler instability suppresses the parallel heat flux similar to the pure parallel driving but there is indication of perpendicular heat flux when compared to the fiducial (only parallel driving) perpendicular fluxes. (b) {\color{black} The spectra of magnetic field fluctuations in the parallel and perpendicular direction as shown in twin y-axis. The perpendicular spectra follows the heat flux driven whistler spectra as expected while the parallel fluctuation spectra appears to have a broken power-law unlike the fiducial case (solid and dashed purple lines).} }
    \label{fig:misaligned}
\end{figure}
\section{Implementation of the force in Boris pusher in the non-relativistic case}
\label{app2}
We use a Boris pusher with normalized velocity, time, and length using $v_{{\rm th}e}$, $\omega^{-1}_{{\rm c}e}$, and $\rho_{ e}$, {\color{black} which are the thermal speed, gyro frequency, and Larmor radius of electrons, respectively}. The force associated with the thermal gradient is independent of the direction of motion and primarily depends on the magnitude of the velocity of the particle. Hence, this is added like an effective electric field (although in our problem there is no {\color{black} imposed} electric field).
\ba
\boldsymbol{E}_{\rm eff} = \boldsymbol{E} + \frac{m_{s}}{q_s}\frac{\hat{\boldsymbol{a}}}{L_{\rm T}}\left(v^2 - \frac{3}{2} v_{{\rm th}s}^2\right).
\ea
The {\color{black} bulk-velocity gradient} force is added at the end of each Boris step by a first order implicit operator-splitting step such that the algorithm is modified as follows (where two consecutive timesteps are indicated, as in the main text, by ${t}^{i-1}$ and ${t}^{i}$):
\color{black}
\ba
\nonumber
\boldsymbol{v}^{-} &=& \boldsymbol{v}^{i-1} + \frac{q_s}{2 m_{s}} \boldsymbol{E}_{{\rm eff}} ({|\boldsymbol{v}^{i-1}|}^2) dt,\\
\nonumber
\boldsymbol{v}^{\prime} &=& \boldsymbol{v}^{-} + {\boldsymbol{v}^{-} \times \boldsymbol{\upsilon}},\\
\nonumber
\boldsymbol{v}^{+} &=& \boldsymbol{v}^{-} + (\boldsymbol{v}^{\prime} \times \boldsymbol{s}),\\
\nonumber
\boldsymbol{v}^{i} &=& \boldsymbol{v}^{+} +  \frac{q_s}{2 m_s} \boldsymbol{E}_{{\rm eff}} ({|\boldsymbol{v}^{+}|}^2) dt, \\
\boldsymbol{v}^{i}_{\rm final} &=& {\boldsymbol{\mathcal{J}}}^{-1} \boldsymbol{v}^{i}.
\ea
where $\boldsymbol{\upsilon} = {q_s}\boldsymbol{B} dt/{2m_s}$, $\boldsymbol{s} = {2 \boldsymbol{\upsilon}}/{(1+|\boldsymbol{\upsilon}|^2)}$, and $\boldsymbol{\mathcal{J}}$ is again defined by (\ref{Jmat_def}) in the main text. As before, the last equation implies a matrix inversion such that $\boldsymbol{\mathcal{J}}$ is non-singular or in other words, it is the solution to a matrix algebraic equation. \color{black}

The effect of these forces on the trajectory of a particle can be tested against the analytic solutions in test cases, e.g., $\boldsymbol{\hat{a}}$ aligned along the guide magnetic field and a shear/compression in the perpendicular $\boldsymbol{\hat{x}}-\boldsymbol{\hat{y}}$ plane. The first force causes the parallel velocity to evolve according to
\ba
\frac{d {v}_{\parallel}}{d{t}} &=& \left({v}^2_{\parallel} +{v}^2_{\perp} - \frac{3}{2} v_{{\rm th}e}^2\right)\frac{1}{{L_{\rm T}}} =\frac{{v}^2_{\parallel} +{v}^2_{\rm r}}{L_{\rm T}}, 
\label{boris_tg}
\ea
which has the following solutions in the special case that $v_{\|}(t = 0) = 0$:
\begin{equation}
\nonumber
{v}_{\parallel} = 
\begin{cases}
{v}_{\rm r} \tan \left(\frac{{v}_{\rm r} {t}}{2{L_{\rm T}}}\right) , & {v}^2_{\rm r}>0 ,  \\
{v}_{\rm r} \tanh \left(\frac{|{v}_{\rm r}| {t}}{2{L_{\rm T}}}\right), & {v}^2_{\rm r}<0 . 
\end{cases}
\qquad . 
\end{equation}
In the above solutions, only the second is bound at all times; the first solution provides a limit on the characteristic timescale ${L_{\rm T}}/{v}_{\rm r}$ for the orbits to be unbounded in the first case {\color{black}(we refer to this condition in our PIC simulations as ``runaway")}. 
In single-particle Boris pusher simulation, we use a static magnetic field $q_e^{-1}\omega_{{\rm c}e}\hat{\boldsymbol{z}}$ (we take $m_e=q_e=1$) and evolve for $t_{\rm stop}=50\omega^{-1}_{{\rm c}e}$. The initial velocity is $v_{{\rm th}e}\hat{\boldsymbol{y}}$ (in this context $v_{{\rm th}e}$ is a parameter rather than having a physical meaning as discussed in section \ref{sec:singpart_tests}) and the temperature gradient scale used in the TF is $L_{\rm T}=5000\rho_e$. The evolution of velocity in the parallel direction and the drift due to TF are shown in the left panel of Fig.~\ref{fig:boris_velo}. We find a very close match with the analytical solution.

The trajectory of a particle due to the {\color{black} bulk-velocity gradient} force can be calculated from the following eigenequation:
\ba
\nonumber
\frac{d{v}_i}{d{t}} &=& 0.5 \mathcal{W}_{ij} {v}_j + \epsilon_{ijk} {v}_j {B}_k,\\
&=& (0.5 \mathcal{W}_{ij} + \mathcal{B}_{ij}) {v}_j = \lambda_i v_i.
\ea
where $\mathcal{B}_{ij} = \epsilon_{ijk} B_k$ is a skew-symmetric matrix with relevant elements given by components of $\boldsymbol{B}$ and $\lambda_j$ denotes the $j^{th}$ eigenvalue of $(0.5 \mathcal{W}_{ij} + \mathcal{B}_{ij})$. Any skew-symmetric matrix has imaginary eigenvalues and accordingly $\boldsymbol{v}_{\perp}$ rotates in the plane perpendicular to the magnetic field. TF contributes a growing part (real eigenvalue) and the orbits should expand gradually. In the single particle Boris pusher simulation, we use a compression/expansion timescale ${\color{black} \tau_{\rm comp}}=200\omega^{-1}_{{\rm c}e}$. We see a very close match of the trajectory of the particle ($\boldsymbol{v}_{y}$) with the analytic solution (right panel of Fig.~\ref{fig:boris_velo}).
\section{Analytical solutions to single particle trajectory in relativistic case}
\label{app3}
In this section, we show the analytical solution to the single particle trajectory for temperature-gradient driven force in the relativistic case (\ref{eq:sing_TG}). To derive this, {\color{black} we use electron's parallel momentum as} $p^2_{\parallel} = m_s^2{c}^2 (\gamma^2_{\rm p} - 1) - p^2_{\perp}$ and similar to the non-relativistic case, the trajectory of the particle depends on the initial $p_{\perp}$.  
\ba
\nonumber
 \frac{d p_{\parallel}}{d{t}} = -\frac{m^2_s{c}^2 \gamma_{\rm p}}{\sqrt{m_s^2{c}^2 (\gamma^2_{\rm p} - 1) - p^2_{\perp}}} \frac{d \gamma_{\rm p}}{d{t}} = \frac{\theta_s c^2}{L_{\rm T}} \Big( \frac{\gamma_{\rm p} - 1}{\theta_s} - \frac{3}{2}\Big).
\ea
The solution to the above equation is of the following form:
\ba
\nonumber
\mathcal{F}_2(\gamma_{\rm p}) - \mathcal{F}_2(\gamma_{\rm p}(t=0)) &=& \frac{\theta_s {c}^2}{L_{\rm T}} {t} .
\ea
where, using $\gamma_{{\rm p}\perp} = \sqrt{1+\gamma^2_{\rm p}v^2_{\perp}}$, $\mathcal{F}_2(\gamma_{\rm p})$ can be expressed as the following:
\ba
\nonumber
&\mathcal{F}_2(\gamma_{\rm p})= -
\theta_s m_s{c} \ln \left( \frac{\gamma_{\rm p}}{\gamma_{\rm p \perp}} + \sqrt{\frac{\gamma^2_{\rm p}}{\gamma^2_{\rm p \perp}} - 1}\right)\\
&+\frac{m_s{c}(1 + \frac{3}{2}\theta_s)}{\sqrt{\frac{9}{4} + \frac{3}{\theta_s} - \frac{p^2_{\perp}}{m^2_s{c}^2 \theta_s^2}}}\ln \left[\frac{\sqrt{\frac{(\theta_s^{-1} + \frac{3}{2}) - \theta_s^{-1}\gamma_{\rm p \perp}}{(\theta_s^{-1} + \frac{3}{2}) + \theta_s^{-1}\gamma_{\rm p \perp}}} + \sqrt{\frac{\gamma_{\rm p}-\gamma_{\rm p \perp}}{\gamma_{\rm p}+\gamma_{\rm p \perp}}}}{\sqrt{\frac{(\theta_s^{-1} + \frac{3}{2}) - \theta_s^{-1}\gamma_{\rm p \perp}}{(\theta_s^{-1} + \frac{3}{2}) + \theta_s^{-1}\gamma_{\rm p \perp}}}-\sqrt{\frac{\gamma_{\rm p}-\gamma_{\rm p \perp}}{\gamma_{\rm p}+\gamma_{\rm p \perp}}}}\right]. \quad
\label{vaytg}
\ea
\section{Wave-particle resonance for relativistic electrons}
\label{sec:resonance}
Here we describe briefly the differences in resonance conditions between non-relativistic and relativistic electrons and the electromagnetic waves. This is important for the discussion of the PIC simulations in section \ref{sec:sims}, where we model two electron-scale microinstabilities which interact with the electrons. 

The first (or any) cyclotron resonance condition {\color{black} for an electron and electromagnetic wave of parallel wave number $k_{\parallel}$} is,
\ba
\nonumber
\omega - k_{\parallel}v_{\parallel} &=& \frac{m\omega_{{\rm c}e}}{\gamma_{\rm p}},\\
\label{eq:reso_gen}
 \implies \omega - \frac{{k}^\prime_{\parallel}p_{\parallel}}{\sqrt{1+\frac{p^2}{m^2_ec^2}}} &=& \frac{m\omega_{{\rm c}e}}{\sqrt{1+\frac{p^2}{m^2_ec^2}}}.
\ea
{\color{black} We write the first equation above in terms of particle's momentum (a convention used in PIC simulations) in the second equation}. We use ${k}_{\parallel}^{\prime}=k_{\parallel}c/m_e$ and $m$ indicates an integer and signifies general cyclotron resonance. In the non-relativistic case, for modes with $k_{\parallel}\approx0$ there is no resonant interaction unless the electron gyration frequency is identical to the wave frequency. But in the relativistic case (taking $m_e=c=1$),
\ba
\label{eq:reso}
p^2_{\parallel} = \left(\frac{m^2\omega^2_{ ce}}{\omega^2} - 1\right) - p^2_{\perp}.
\ea
The above is circular like the black band (semi-circular) in momentum space that we see in our heat flux anisotropy and temperature anisotropy at early times (Fig.~\ref{fig:hwianisotropy} and Fig.~\ref{fig:presanisotropy}) and there is no resonance with $m=0$. For non-zero $k_{\parallel}$, (\ref{eq:reso_gen}) can be written in a compact form in terms of the index of refraction $\boldsymbol{n}_{\parallel} = \boldsymbol{k}_{\parallel}c/\omega$ and $l_{\rm m} = m\omega_{{\rm c}e}/\omega$ as:
\ba
\label{eq:reso_gen2}
p^2_{\parallel}(n^2_{\parallel} - 1) + 2n_{\parallel} l_{\rm m} p_{\parallel} -p^2_{\perp} + (l^2_{\rm m} - 1)=0
\ea
For $|n_{\parallel}|=1$, the above represents a parabola with no resonance at $m=0$. The case for $n_{\parallel}\neq 1$ can be most generally written as following with elliptical/hyperbolic momenta space curves:
\ba
\label{eq:reso_gen3}
\frac{{\Big[p_{\parallel}+{n_{\parallel}l_{\rm m}}/{(n^2_{\parallel} -1)}\Big]}^2}{{(n^2_{\parallel}-1)}^{-1}} - p^2_{\perp} = 1 - l^2_{\rm m} + \frac{n^2_{\parallel}l^2_{\rm m}}{n^2_{\parallel}-1} . 
\ea
It is unlikely that $|n_{\parallel}|<1$, as this would be a violation of speed of light threshold. For $m=0$, the hyperbolas are evenly spaced in the parallel axis. Generally, to obtain resonance uniformly along the entire $p_{\parallel}$ axis, oblique waves or left circularly polarised parallel waves are necessary. The center of the hyperbolas defined by above ($m\neq0$) is shifted along the negative $p_{\parallel}$ for a wave which propagates along the field. These effects may be demonstrated in the momenta space ($p_{\parallel}-p_{\perp}$) electron distribution map as bending dark lines of resonance at early times. 
\section{Parallel heat flux scaling with temperature and plasma $\beta_s$}
\subsection{Scaling with temperature}
\label{app4:theta}
The parallel heat flux can be calculated in the {\color{black} weakly} relativistic case, considering that the energy carried by particles within $\mathrm{d}p_{\parallel} \mathrm{d}p_{\perp}$ is, 
\ba
\epsilon=m_sc^2\sqrt{1 + \frac{p^2}{m_s^2 c^2}}=m_sc^2\left(1 + \frac{p^2}{2m_s^2 c^2}\right),
\ea
and using assumptions of small $p$ and $m_s=c=1$, 
\ba
\nonumber
\dot{q}_{\parallel} &=& \int^{\infty}_{0}\int^{\infty}_{-\infty} \mathrm{d}p_{\parallel} \mathrm{d}p_{\perp} \; 2\pi p_{\perp} p_{\parallel} \frac{\mathrm{d}f}{\mathrm{d}t} (\epsilon -1)  \\
\nonumber
&=& \int^{\infty}_{0}\int^{\infty}_{-\infty} \mathrm{d}p_{\parallel} \mathrm{d}p_{\perp} \, 2\pi p_{\perp} p_{\parallel} S_{\rm p} (\epsilon -1)\\
\nonumber
&=& \frac{1}{L_{\rm T}} \int^{\infty}_{0}\int^{\infty}_{-\infty} \mathrm{d}p_{\parallel} \mathrm{d}p_{\perp} \; 2\pi p_{\perp} p_{\parallel} \gamma_{\rm p}^{-1}\frac{p^2}{2}  \\
\nonumber
&\times& \left[p_{\parallel}\Big(\frac{\gamma_{\rm p} -1}{\theta_s} - \frac{5}{2}\Big)\right]\frac{e^{-\frac{\sqrt{1+p^2}}{\theta_e}}}{4\pi \theta_s K_2({\theta_s^{-1}})} \\ 
&\approx& \frac{5\sqrt{2\pi}{\theta_s}^{5/2} e^{-{\theta_s}^{-1}}}{4 L_{\rm T}K_2({\theta_s^{-1}})}.
\label{eq:qpar}
\ea
The analytical heat flux matches well for multiple $\theta_e$. In this section, we provide an example in Fig.~\ref{fig:htflx_theta} that includes the entire momentum phase space scanned by the electrons in our PIC simulations to match the above heat flux reasonably well.

\subsection{Scaling with $\beta_e$}
\label{app4:beta}
The parallel heat flux is expected to be suppressed as inverse of $\beta_e$ {\color{black} (relative to the free-streaming value)} at the saturated stage of the heat flux driven whistler instability. This occurs if the physical size of the PIC box we simulate allows for sufficiently large wavelengths of the instability to resonantly scatter suprathermal electrons. In section \ref{sec:alignedgrad}, we discuss this scaling by calculating the parallel heat flux with limits on the maximum $p/m_ec$ that can be scattered by the whistlers. Here we discuss the net parallel heat flux in the PIC box along with potential contributions by runaway electrons due to unbounded orbits (discussed in section \ref{sec:singpart_tests}). 

For our simulations with $\beta_e \in [20,40,60]$, TF can dynamically kick a small fraction of electrons to a significantly large parallel momentum. This causes the small fraction to carry high energy along the magnetic field. Unless these are scattered by self-generated long-wavelength whistlers or non-resonant scattering by the microinstability or particle noise in our simulations, these can cause anomalous rise in heat flux. The heat flux regulated by whistlers or non-resonant scattering should scale with inverse of $\beta_e$ while that regulated by noise can contribute a constant background parallel heat flux. In Fig.~\ref{fig:htflx_all}, we show the net heat flux (left panel), and saturated heat flux (right panel). In the former, we compare the net heat flux between small and large box for $\beta_e=40$ (solid purple and dashed black lines). The large box allows for longer wavelength whistlers and hence the regulation is more efficient such that the parallel heat flux scales well as $1.6\beta^{-1}_e$ after we subtract a background heat flux $q_{\rm bg}=0.045{(2\theta_e)}^{\frac{3}{2}}$. However, the simulation with $\beta_e=60$ fails to scale as well since it is carried out in a box that allows a maximum wavelength that scatters electrons with maximum momentum $\left|p/m_ec \right|\sim 3.28\sqrt{(2\theta_e)}$. In non-relativistic simulations of PIC box with hot and cold reservoirs along the magnetic field, it is usually found that the saturated heat flux has contributions from regulated suprathermal electrons of velocities $\sim 4v_{{\rm th}e}$ (which is slightly higher than what we get for $\beta_e=60$ simulation due to the PIC box size). On the other hand, when we limit the integral over the momenta range that can be scattered by whistlers, we find a previously known scaling $1.5\beta^{-1}_e$ (middle panel in Fig.~\ref{fig:time_evolve} and see \cite{2018JPP_komarov}). The right panel in Fig.~\ref{fig:htflx_all} also clearly demonstrates a good fit to the net parallel heat flux to saturate at $\sim 1.6 \beta^{-1}_e$. 

A potential contribution to mildly enhanced heat flux -- indicated by pre-factor $1.6$ as opposed to pre-factor $1.5$ that we get when we limit the heat flux calculation to the range of electron momenta that can be scattered-- can, in principle, come from runaway electrons regulated by noise. In the left panel of Fig.~\ref{fig:htflx_all}, we show the total contribution from the fraction of electrons that dynamically cross the orbital runaway condition (dot-dashed lines). It is evident that in $\beta_e=20$, the saturated heat flux may have a small contribution from such electrons. However, it is also regulated by particle noise (as is evident in our simulations by the $\sim 10\%$ increase in $\theta_e$ with time and that our shot noise should scale as $1/\sqrt{N_{\rm ppc}} = 0.02$ where $N_{\rm ppc}$ is the particle-per-cell). Such a contribution is reflected in our fitted curve with $q_{\rm bg}=0.045{(2\theta_e)}^{\frac{3}{2}}$. This contribution from noise-regulated relativistic electrons and possibly inefficient non-resonant scattering is not required to be considered in the pure whistler-regulated regime (middle panel in Fig. \ref{fig:time_evolve}).
\begin{figure}
    \includegraphics[width=8.cm]{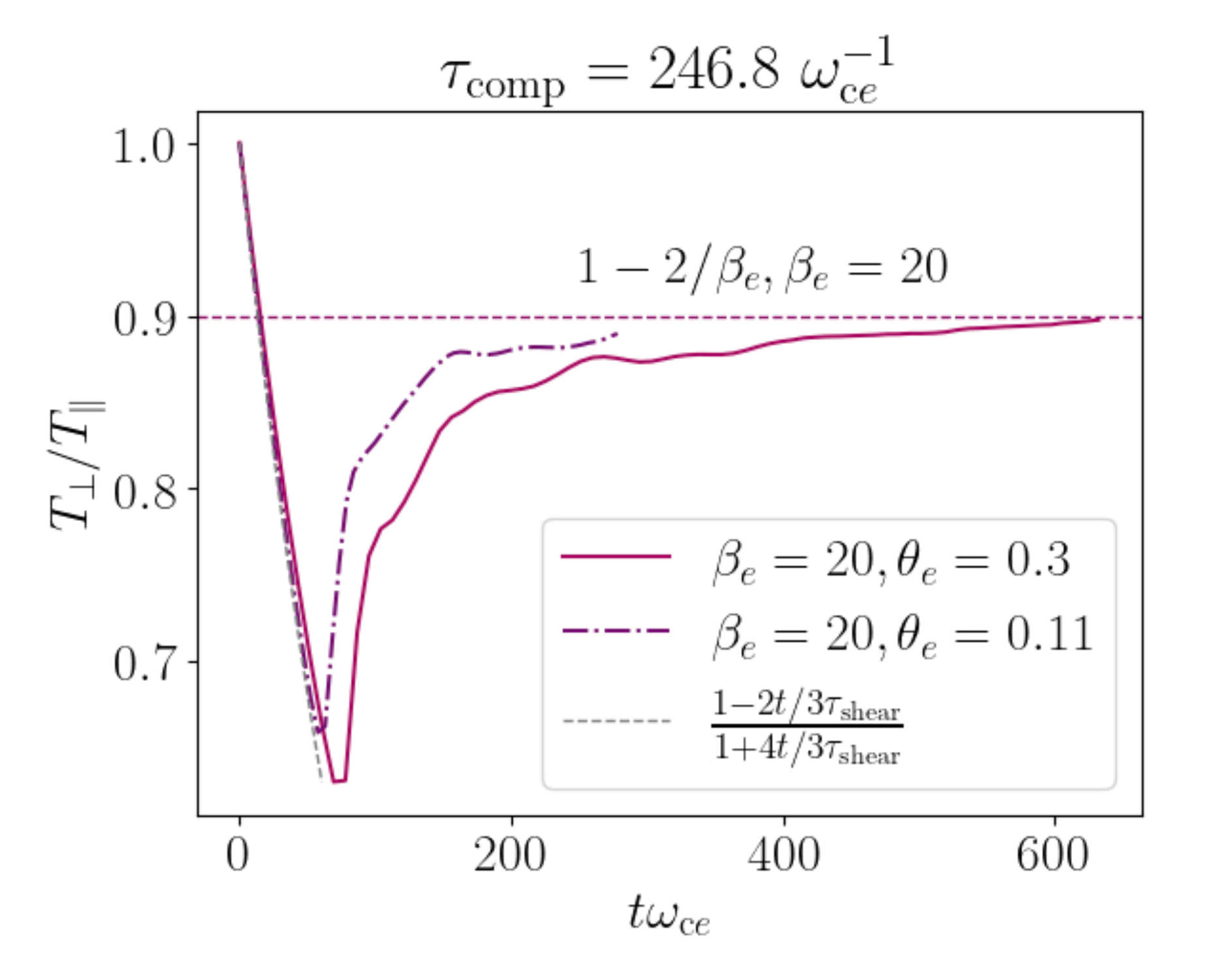}
    \centering
    \caption{Comparison of the temperature anisotropy with time between simulations of $\theta_e\in[0.11,0.3]$(see Appendix \ref{app6}). }
    \label{fig:shear_th}
\end{figure}
\section{Misaligned heat flux and magnetic field}
\label{app5}
In this section, we include the simulation with $\beta_e=60$ (Fig.~\ref{fig:misaligned} top panel) and the force driven by the temperature gradient along $\boldsymbol{\hat{x}}$ and $\boldsymbol{\hat{y}}$ so that along the former, the heat flux grows similarly to the case with only parallel driving (net force at $\alpha=\pi/4$ angle from either axes and smaller $L_{\rm T}$ by a factor of $\cos\alpha$). The top panel shows parallel and perpendicular heat fluxes. While the former shows signature of whistler regulation as expected, the latter is not identical to the corresponding perpendicular flux in the fiducial case with a field aligned TF (orange solid/black dotted lines and brown solid/blue dashed lines). The variations are below the noise level, but the shot noise must be identical in both simulations. Thus, the orange line may imply an enhanced flux in perpendicular out-of-plane direction. The increase beyond $400\omega^{-1}_{{\rm c}e}$ can be driven by runaway electrons regulated by noise since it is present in both simulations. We need to carry out this exploration with higher particle-per-cell ($N_{\rm ppc}$) in future, specifically with the physical parameters in the regime $\rho_e\beta_e/L_{\rm T}\sim 1$ to allow the possibility of larger perpendicular fluxes. We present the spectra of magnetic field fluctuations in the parallel and perpendicular direction in the bottom panel of Fig.~\ref{fig:misaligned}. The perpendicular spectra is following the heat flux driven whistler spectra, while there is difference in the power carried by parallel field fluctuations between the two simulations. 

The simulation with misaligned field, closer to the realistic turbulent field morphology, indicates that the accepted anisotropic thermal conduction model may or may not hold. An exploration along this direction has been beyond the scope of PIC simulations before our proposed TF-PIC method.

\section{Analytical estimate of the growth of pressure anisotropy}
\label{app6}
The growth of temperature anisotropy due to TF ({\color{black} bulk-velocity gradient}) can be calculated from the source term as the following at $t\ll t_{\rm growth}$ where $t_{\rm growth}\sim 20~\omega^{-1}_{{\rm c}e}$ is the linear growth timescale of firehose:
\ba
\nonumber
T_{\parallel} &=& \int^{\infty}_{0}\int^{\infty}_{-\infty} \mathrm{d}p_{\parallel} \mathrm{d}p_{\perp} \, 2\pi p_{\perp} p^2_{\parallel, \perp} \gamma^{-1}_{\rm p} (f-f_0)\\
\nonumber
&=&  \int^{\infty}_{0}\int^{\infty}_{-\infty} \mathrm{d}p_{\parallel} \mathrm{d}p_{\perp} \, 2\pi p_{\perp} p^2_{\parallel, \perp} \gamma^{-1}_{\rm p} S_{\rm p} t \nonumber \\
\nonumber
&=& \frac{1}{\theta_e {\color{black} \tau_{\rm comp}}}\int^{\infty}_{0}\int^{\infty}_{-\infty} \mathrm{d}p_{\parallel} \mathrm{d}p_{\perp} \; 2\pi p_{\perp} p^2_{\parallel, \perp} \gamma^{-2}_{\rm p}  \\
&& \qquad \qquad \qquad  \times \left(\frac{4}{3} p^2_{\parallel} - \frac{2}{3}p^2_{\perp}\right) t f_{{\rm M}s} , \\
T_{\perp} 
&=& \frac{1}{2\theta_e {\color{black} \tau_{\rm comp}}}\int^{\infty}_{0}\int^{\infty}_{-\infty} \mathrm{d}p_{\parallel} \mathrm{d}p_{\perp}\; 2\pi p_{\perp} p^2_{\parallel, \perp} \gamma^{-2}_{\rm p} \quad \nonumber \\
&& \qquad \qquad \qquad \times \left(\frac{4}{3} p^2_{\parallel} - \frac{2}{3}p^2_{\perp}\right) t f_{{\rm M}s} ,
\ea
where the pre-factor $1/2$ is included for perpendicular temperature only.
\ba
\nonumber
T_{\parallel} &\approx& \frac{4 t \theta_e}{3 {\color{black} \tau_{\rm comp}}}, \\
T_{\perp} & \approx& -\frac{2 t \theta_e}{3 {\color{black} \tau_{\rm comp}}}.
\ea
We ignore some higher order terms in $\theta_e$ in the above. The temperature anisotropy is as follows,
\ba
\frac{T_{\perp}+T_{s}}{T_{\parallel}+T_{s}} = \frac{\left[{1-{2 t}/{3 {\color{black} \tau_{\rm comp}}}}\right]}{\left[{1+{4 t}/{3 {\color{black} \tau_{\rm comp}}}}\right]} \approx \left(1- \frac{2t}{{\color{black} \tau_{\rm comp}}}\right),
\ea
where $T_s = m_ec^2\theta_e$ is the isotropic temperature and $m_e=c=1$ is assumed for the calculation aligned with the PIC simulations. Fig.~\ref{fig:shear_th} shows the time evolution of the anisotropy with different $\theta_e$. The source term implies that in the phase space, the amplitude of the anisotropy is inversely dependent on $\theta_e$ and hence the nature of the firehose modes may change. But the net anisotropy is approximately independent of the temperature.

\bibliography{apssamp}

\end{document}